\documentclass[twocolumn]{aastex631}

\newcommand{\D}{\mathrm{d}}
\newcommand{\Ms}{{\ensuremath{\mathrm{M}_{\sun}}}}

\newcommand{\dens}{{\ensuremath{\mathrm{cm}^{-3}}}}
\newcommand{\K}{{\ensuremath{\mathrm{K}}}}
\newcommand{\Myr}{{\ensuremath{\mathrm{Myr}}}}
\newcommand{\Mpc}{{\ensuremath{\mathrm{Mpc}}}}
\newcommand{\FeH}{\ensuremath{[\mathrm{Fe}/\mathrm{H}]}}

\submitjournal{ApJ}

\shorttitle{PISN metal mixing}
\shortauthors{Magg et al.}

\begin{document}

\title{Metal Mixing in Minihalos: The Descendants of Pair-Instability Supernovae}

\correspondingauthor{Mattis Magg}
\email{mattis.magg@protonmail.com}

\author[0000-0002-9022-5136]{Mattis Magg}
\affiliation{Institut für Theoretische Astrophysik, Albert-Ueberle-Str 2, 69120, Heidelberg, Germany}

\author{Anna T. P. Schauer}
\altaffiliation{Hubble Fellow}
\affiliation{Department of Astronomy, University of Texas at Austin, Austin, TX 78712, USA}

\author{Ralf S. Klessen}
\affiliation{Institut für Theoretische Astrophysik, Albert-Ueberle-Str 2, 69120, Heidelberg, Germany}
\affiliation{Universität Heidelberg, Interdiszipliäres Zentrum für Wissenschaftliches Rechnen, Im Neuenheimer Feld 205, 69120 Heidelberg, Germany}

\author{Simon C. O. Glover}
\affiliation{Institut für Theoretische Astrophysik, Albert-Ueberle-Str 2, 69120, Heidelberg, Germany}

\author{Robin G. Tress}
\affiliation{Institut für Theoretische Astrophysik, Albert-Ueberle-Str 2, 69120, Heidelberg, Germany}

\author{Ondrej Jaura}
\affiliation{Institut für Theoretische Astrophysik, Albert-Ueberle-Str 2, 69120, Heidelberg, Germany}



\begin{abstract}
The lack of observations of abundance patterns originating in pair-instability supernovae has been a long-standing problem in relation to the first stars. This class of supernovae is expected to have an abundance pattern with a strong odd-even effect, making it substantially different from present-day supernovae. In this study, we use a cosmological radiation hydrodynamics simulation to model such supernovae and the subsequent formation of the second generation of stars. We incorporate streaming velocities for the first time. There are 14 star-forming minihalos in our 1\,cMpc\,$h^{-1}$ box, leading to 14 supernovae occurring before redshift $z=19.5$, where we start reducing the complexity of the simulation. Following the explosions, extremely metal-poor stars form in 10 halos via internal and external enrichment, which makes it the most common outcome. Only one halo does not recollapse during the simulations. This result is at tension with the current (lack of) observations of metal-poor stars with pair instability supernova abundance patterns, suggesting that these very massive stars might be rare even in the early Universe. The results from this simulation also give us insights into what drives different modes of recollapse and what determines the mixing behavior of metals after very energetic supernovae.
\end{abstract}


\section{Introduction}\label{sec:intro}
The first stars, labelled Population~III (Pop~III) stars, form a few hundred million years after the Big Bang \citep{GloverReview, HaemmerleReview}. They mark the end of the cosmic dark ages and synthesize the first heavy elements. Yet they have not been observed directly thus far and this is unlikely to change soon \citep{Schauer2020}, as even the next generation of telescopes will not be able to capture the tiny galaxies in which they form \citep{Jeon2019}. Theoretical models predict that, due to the less efficient cooling of metal-free gas, Pop~III stars should on average be more massive than the average star in the present day Universe \citep{BrommReview, GreifReview}. However, their initial mass function (IMF) remains poorly constrained and there are large differences between the results of different simulations \citep{Clark11, Greif11b, Hirano15, Hosokawa16, Stacy16, Susa2019}.

In the absence of direct observations or convergence of theoretical models, there remain three commonly used methods to improve our understanding of Pop~III stars. Firstly, the non-detection of metal-free stars in the Milky Way can be used to infer limits on the formation of long-lived, low-mass Pop~III stars \citep{Oey2003, Tumlinson2006, Salvadori2007, Hartwig15b, Ishiyama16, Magg18, Magg19, Rossi2021}. However, this method only allows us to constrain Pop~III star formation in the mass range below 1\,\Ms, because only such low-mass stars have long enough lifetimes to survive until today \citep{Marigo2001}.
 
The second method involves constraining the formation of the first stars with the 21\,cm absorption feature at high redshifts \citep{Pritchard2010, Fialkov14, Fialkov19}. This method may give insight into the integrated properties of Pop~III stellar populations, but it will be difficult to determine details about individual Pop~III stars with this method. 

The third and most used method is stellar archaeology \citep{Frebel15, Fraser17, Ishigaki18}. In this approach, the observed abundance patterns of extremely metal-poor (EMP) stars, i.e., stars with an iron abundance\footnote{For elemental abundances, we use the standard notation \\$\mbox{[X/H]} = \log_{10}(N_\mathrm{X}/N_\mathrm{H})-\log_{10} (N_{\mathrm{X},\sun} /N_{\mathrm{H},\sun})$ where $N_{\mathrm{X}}$ and $N_{\mathrm{H}}$ are the fractional abundances by number of any element X and hydrogen, and $N_{\mathrm{X},\sun}$ and $N_{\mathrm{H},\sun}$ are the corresponding solar abundances.} of less than $\mbox{[Fe/H]} = -3$, are matched to the abundance patterns found in models of Pop~III supernovae (SNe), such as the models of \citet{HegerWoosley2010} or \citet{Ishigaki18}. It is usually assumed that exactly one SN contributes to the metal abundances in each EMP star, and that the SN model which matches the observed abundance pattern best is the most likely progenitor of the observed star \citep[for a critical assessment of this approach see][]{Hartwig18a, Magg20}.

The question of pair-instability SNe (PISNe) has been a long-standing problem in this field. Very massive metal-free stars (between around 140 and 260\,\Ms) are predicted to undergo this very energetic type of SN \citep{Bond84} in its oxygen core phase. For example, \citet{HegerWoosley2002} predict explosion energies between 10 and 100 foe\footnote{1\,foe is $10^{51}\,\mathrm{erg}$; the abbreviation stands for (ten to the) Fifty-One Erg} and very large metal yields, up to 60\,\Ms\ of iron. Such theoretical models also result in a very strong odd-even effect, i.e., a very large difference in the abundances between elements with odd and with even atomic numbers. However, to this day and amongst hundreds of observed metal-poor stars, only one candidate has been found for a star that may have been (partially) enriched by a PISN \citep{Aoki14}. This is surprising, considering that PISNe are a stable and long-standing prediction of stellar evolution models \citep{HegerWoosley2002, HegerWoosley2010, Nomoto13}. We also know of the existence of stars in this mass-range in the present-day Universe \citep[see e.g.][]{Bestenlehner2020}, and despite large differences in their predictions, simulations of star formation agree that massive stars become more common at low metallicities, not less \citep{bsmith15, Chiaki17}.

There have been two main hypotheses to explain the lack of observational signatures of PISNe while still assuming that these SNe do occur and are not exceedingly uncommon or significantly different from what is predicted. Firstly, \citet{Whalen08b} found in one-dimensional simulations that the host halos of PISNe are almost completely cleared of gas due to the very energetic explosions. This prevents further star formation from occurring within them. Secondly, \citet{Karlsson2008}, \citet{deBennassuti17}, and \citet{Salvadori2019} suggest that because PISNe produce such a large amount of metals, their descendants should primarily not be found among the EMP stars, but at higher metallicities. This would explain the lack of detections because there are many more stars at higher metallicities in the Galaxy \citep[e.g.][]{Youakim2020}, and because there are fewer dedicated efforts to determine detailed abundance patterns of thgese objects. If this hypothesis is correct, both factors would decrease the likelihood of finding the descendants of PISNe.

Here, we use a 3D cosmological hydrodynamical simulation run with the code \textsc{arepo} \citep{arepo} in order to investigate when the descendants of hypothetical PISNe are formed and what metallicities they have. For the first time, we include the effects of baryonic streaming velocities into this type of metal-tracing simulation.

We explain our methodology and present information about the simulation set-up and resolution in Section \ref{sec:methods}. Results of the simulations, especially the metallicity of the second-generation stars and the time of their formation are shown in Section \ref{sec:results}. We discuss the implications of these results in Section \ref{sec:discussion}, and summarize our findings in Section \ref{sec:sum}. To distinguish between comoving and physical length-units we use the following convention: comoving units are always scaled by the Hubble parameter $h$, i.e., the Hubble constant $H_0$ in units of $100\,\mathrm{km}\,\mathrm{s}^{-1}\,\Mpc^{-1}$. For example, the box length of the simulation -- 1 $\Mpc\,h^{-1}$ -- is a comoving length, while physical length-scales are given without the dependence on $h$.

\section{Methods} \label{sec:methods}
\subsection{Overview}
In order to model PISNe in an appropriate environment, a complex multi-physics simulation setup is necessary. We need to self-consistently treat the formation of the first stars in minihalos in a cosmological context, the radiative feedback which shapes the environment in which the SNe explode, the SNe themselves and then we need to follow the metals until a second generation of stars forms.

Previous simulations of this type have focused on one single or a few selected halos \citep[e.g.][]{Greif10, Jeon14, Ritter15, bsmith15, Chiaki18}. They model the formation of one or a few stars of a pre-selected mass in one or up to three \citep[in][]{Chiaki18} minihalos, and include the effects of radiative feedback, supernovae and the formation of the first enriched stars. As the resulting mixing processes in these studies show large variations, it is necessary to simulate a large region that includes a statistically significant sample of halos to gain a more complete picture of the variations in outcomes. Additionally, in order to observe the recollapse of the halos, the simulation needs to be run for an extended time-period. For example, \citet{Latif2020} simulate PISNe in five different halos and find that only the two most massive halos ($M_\mathrm{vir}>10^7\,\Ms$) form stars again within 31\,\Myr\ after the SNe, demonstrating that in order to determine the ultimate fate of the gas enriched by PISNe, we need to follow its evolution for tens of Myr.

\subsection{Setup and initial conditions}
In order to have as many star-forming halos as possible for our study, we simulate the largest box we can afford at the resolution required to properly resolve gas cooling and star formation (see Section \ref{sec:ref}) in minihalos. We use a periodic cosmological box with an edge-length of 1\,Mpc\,$h^{-1}$. The simulation is initialized at $z=200$ and the initial conditions are created with \textsc{MUSIC} \citep{music}. We use cosmological parameters from \citet{Planck2015} which are $\Omega_\mathrm{b} = 0.04864$, $\Omega_\mathrm{m} = 0.3089$, $\Omega_\Lambda = 0.6911$, $\sigma_8=0.8159$, $n_\mathrm{s} = 0.961$ and $H_0= h\,100\mathrm{km}\,\mathrm{s}^{-1}\,\mathrm{Mpc}^{-1}$ with $h = 0.6774$. We generate $512^3$ dark matter particles and equally many gas cells, giving us a resolution of approximately 800\,\Ms\ for the dark matter particles and an initial mass of 150\,\Ms\ for the gas cells. \textsc{Arepo} ensures that the gas cell masses remain within a factor of two of this initial mass at later times, unless one of the resolution criteria described in Section~\ref{sec:ref} applies. Taking the refinement into account, a typical star forming minihalo (of around $10^6\,\Ms$) consists of 1300 dark matter particles and $5\times 10^5$ gas cells, reaching a mass resolution down to $0.01\,\Ms$ in the highest density gas. The simulation presented first in this study used 3 million core-hours on SuperMucNG at the Leibnitz Computing Centre (LRZ) in Garching.

We approximate the effects of supersonic baryonic streaming \citep{Tseliakhovich10} by adding a velocity offset of 4.9\,km\,s$^{-1}$ in the positive $x$-direction to the baryons in our initial conditions. This value corresponds to 0.8 times the root-mean-square streaming velocity and therefore represents the most likely value \citep{Schauer2019b}. Fully accounting for the effects of supersonic baryon streaming would require changing the density distribution in our initial conditions. However, these effects are only of secondary importance for simulations like ours \citep{Park2020}.

In order to speed up the simulation we deactivate the radiative transfer and Pop~III SNe at $z=19.5$, after the first 14 Pop~III SNe exploded. This is shortly before the first metal-enriched star forms, and as our simulations do not yet have the capability to appropriately model Pop~II stars or identify them at run-time, we would start spending substantial computational resources on modelling these stars as if they were metal-free. The shut-off occurs long before large-scale reionization is expected to affect the simulations notably. Reducing the complexity of the simulation at this point allows us to run the simulation for an extended period and capture more recollapsing halos. Modelling the entire period with full physics would require an implementation of metal-enriched star formation and low-metallicity chemistry, both of which are beyond the scope of this project.

We generate snapshots of the simulation every $\Delta z=0.5$, starting at $z=25$. This corresponds to a time difference between snapshots ranging from 4\,Myr at $z=25$ to 15\,Myr at $z=14$. In order to capture the formation of the second generation stars, we additionally take snapshots at the next full-hydro step\footnote{To satisfy the Courant-Friedrichs-Levi condition \citep{CFL}, smaller cells need to be simulated on smaller time-steps. In order to not waste computational resources, \textsc{Arepo} uses an adaptive time-stepping scheme, in which small cells are simulated on smaller time-steps than larger cells. The largest time-step, i.e., a time-step in which all simulated cells are evolved, is referred to as "full-hydro step"} after each sink particle (see Section \ref{sec:sinks}) forms and at the last full-hydro step before each SN. 

\subsection{Simulation framework}
Our basic setup derives from the cosmological simulations performed by \citet{Schauer2019a, Schauer2021} and is adapted as necessary for our simulation. We model the formation of the first stars, the explosion of the most massive of these stars as SNe and the formation of the second generation of stars with the moving mesh code \textsc{arepo} \citep{arepo}. We include the improved integration schemes from \citet{Pakmor16} and the improved mesh regularization from \citet{Mocz15}. Chemistry and cooling of the gas are modelled with the same treatment as in \citet{Schauer2019a}. We follow the non-equilibrium chemistry of H, D, He, H$_2$, HD and their corresponding ions by using the primordial chemistry network from \citet{Glover08}, including updates introduced by \citet{Glover15}. The effects of metal cooling are not included, as it is not relevant at low metallicity at the densities we reach in our simulation \citep{Jappsen07, bsmith15, Chiaki16}.

\subsection{Star formation and sink particles}
\label{sec:sinks}
To model the formation of stars, we use an approach based on sink particles that was introduced in \citet{Tress2020a}. Sink particles are collisionless particles which are able to absorb gas that is gravtationally unstable and collapsing onto the particle. They are used to represent gravitationally collapsing regions as resolving and simulating such regions in numerical simulations is extremely expensive \citep[see e.g.][]{Bate95, Federrath2010, Greif11b}. The formation of sink particles depends on a density threshold and the radius of the region from which a sink particle can be formed, the so-called sink-formation radius. A cell is turned into a sink particle if all of the following criteria are fulfilled:
\begin{itemize}
    \item the density in the cell is larger than the density threshold;
    \item there are no additional sink particles within one sink-formation radius;
    \item the cell is at a local potential minimum;
    \item the gas within one sink-formation radius around the cell is converging ($\vec{\nabla} \cdot \vec{v} < 0$) and collapsing ($\vec{\nabla}\cdot\vec{a} < 0$);
    \item the sphere of gas around the cell (again, with a size of one sink-formation radius) is gravitationally bound, i.e., its potential energy is larger than twice the sum of its kinetic and thermal energy.
\end{itemize}
We note that the latter criterion is especially important. When it is deactivated, we see many sink particles forming in the strong shocks produced by the SNe, which is a numerical artifact, as this shocked gas is not actually gravitationally unstable.

After they have formed, sink particles can accrete gas. In order for gas to be accreted from a grid cell, the density must be above the threshold density and the cell must be located within the sink-formation radius of the sink. In addition, the gas must be gravitationally bound to the sink particle and collapsing onto it. Once a cell is marked as accreting, gas is removed from it until it reaches the threshold density. The gas mass removed from a cell during a single time-step is additionally limited to 90 per-cent of the cells initial mass. The gas removed from the cell is added to the sink particle, and the momentum of both cell and sink particle are updated as appropriate in order to ensure momentum conservation. ``Skimming'' the mass from the cells in this way helps to avoid grid construction problems and other numerical artifacts that can occur if the entire cell is accreted.

These methods were originally developed for simulations of the present-day Universe and we have adapted them to be compatible with the comoving internal unit-system of \textsc{arepo}. The density threshold for sink particle formation in the simulations presented here is $10^4\,\dens$. We chose this value for two reasons. Firstly, primordial gas is expected to become Jeans-unstable at these densities \citep{yoha06, Greif11b}, and therefore we can be confident that it begins to collapse at this density and will end up forming stars. Secondly, we do not model how the metals injected by the first SNe affect the chemistry and cooling. At the metallicities encountered in our simulation, metal-induced fragmentation is expected to only occur at number densities higher than around $10^4\,\dens$ \citep{bsmith15, Chiaki15}, and so our simulation will not be affected by our neglect of metal cooling and chemistry. Densities higher than this should be considered unresolved in our simulation and further refinement (see Section~\ref{sec:ref}) is deactivated above this threshold. The sink-formation radius is chosen such that the mass within a sphere with a mean density equal to the threshold density is exactly one Jeans mass at a temperature of $T=200\,\K$, which is a typical temperature for the point at which primordial gas becomes Jeans unstable \citep{yoha06}. In practice, this corresponds to a sink-formation radius of 2\,pc.

As we are focusing exclusively on PISNe, we assign a single star with a fixed mass of 200\,\Ms\ to each newly forming sink particle. We note that stars are only assigned to the sink-particle for the modelling of stellar feedback. The gravitational mass of the sink-particle is not altered by this choice. For future studies, we plan to adapt the star formation recipe based on Poisson sampling introduced in \citet{Sormani17} to the case of a Pop~III IMF, but we do not use this in our present work.

\subsection{Radiative transfer}
We use the \textsc{simplex}-based \citep{Kruip10, Paardekooper2010} radiative transfer code \textsc{sprai} \citep{Jaura18, Jaura20} to model the radiative feedback of the stars. Radiation is modelled in four energy bands ranging from 11.2\,eV to 136\,eV, which are detailed in Table \ref{tab:rad_bins}. The photon emission rates in these bands are calculated assuming that the Pop III stars radiate as black bodies, with an effective stellar temperature and radius obtained by interpolation of the data in \citet{Schaerer2002}. We compute the average cross-section for each modelled photochemical process in each energy band by weighting the frequency-dependent cross-sections by a black-body spectrum of temperature $T=80000\,\mathrm{K}$, as described in more detail in \citet{Fervent} and \citet{Jaura18}. We model the non-equilibrium radiation-chemistry, the corresponding heating and the radiation pressure exerted by the absorbed photons. Because \textsc{sprai} can only run on the largest time-steps within the adaptive time-stepping of \textsc{arepo}, we limit the maximum time-step to be around 10000 years at redshifts below $z=25$, i.e., when the first stars form in our simulation. The life-times of the stars are assigned as interpolated values from \citet{Schaerer2002}, which results in a life-time of 2.2\,Myr for a 200\,\Ms\ Pop~III star.

\begin{deluxetable}{llcc}
    \tablecaption{\label{tab:rad_bins}Summary of radiation energy bins}
    \tablecolumns{4}\colnumbers
    \tablehead{ \colhead{Energy (eV)} & \colhead{Processes} & \colhead{$\sigma\, (10^{-18}\, \mathrm{cm}^{2})$} & \colhead{$\dot{N} (10^{48}\mathrm{s}^{-1})$}}
        \startdata
 11.2--13.6 & H$_2$ dissociation & $2.47$ &9.15\\ ~\\
  13.6--15.2& H$_2$ dissociation & $2.47$& 6.15\\
  & H ionization&$5.35$  & \\ ~\\
  15.2--24.6& H$_2$ ionization & $6.52$&32.4\\
  &H ionization &$2.39$  &\\ ~\\
  24.6--136&H$_2$ ionization & $1.85$&47.5\\
  &H ionization & $0.51$&\\
  &He ionization & $4.65$&\\
     \enddata
     \tablecomments{The columns are: (1) Energy bin, (2) Modelled photo-chemical processes, (3) reaction crossection for the processes, and (4) photon emission rate of a single 200\,\Ms\ star in the energy bin.}
\end{deluxetable}

\subsection{Supernovae}
At the end of each star's life, the SN explosion energy is injected as thermal energy into the closest 1000 cells, as in \citet{Tress2020a}. The process is again adapted to the internal cosmological unit system of \textsc{arepo}. We ensure that the Sedov-Taylor phase of each SNe is accurately captured (see Section \ref{sec:ref}), and therefore we do not need to make use of a momentum injection scheme \citep[e.g.][]{Gatto15}. Each SN explodes as a very energetic PISN \citep{HegerWoosley2002} with an explosion energy of $E_\mathrm{SN}=100\,\mathrm{foe}$. In addition to the energy, we also inject $10^7$ Monte-Carlo tracer particles \citep{Genel13} into each SN. We take these tracer particles to represent the metals injected into the SN. They are used to follow the enrichment and to determine the metallicity of the second generation of stars. 

\subsection{Resolution and refinement}\label{sec:ref}
The variety of physical processes modelled in our simulation translates to a set of resolution criteria. Specifically we need to resolve:
\begin{enumerate}
    \item the formation and substructure of minihalos,
    \item the gravitational collapse and fragmentation of molecular clouds,
    \item the expansion of ionized regions around massive stars,
    \item the expansion of SN remnants.
\end{enumerate}
This translates into the following resolution criteria:
\begin{enumerate}
    \item The dark matter particle mass needs to be well below the halo mass. \citet{Schauer2019a} demonstrated that at least 1000 dark matter particles are required to consider a halo resolved. We therefore require a dark matter particle mass of no more than 1000\,\Ms\ in order to resolve a $10^6\,\Ms$\ minihalo. We fulfil this criterion with our dark matter particle mass of 800\,M$_\odot$.
    \item In order to find numerically converged fragmentation behavior, the Jeans length has to be resolved with several resolution elements \citep{Truelove97}. We therefore require that the Jeans length is always resolved with at least 8 cells, leading to a maximum cell mass of
    \begin{equation}
     M_\mathrm{c, Jeans} = \frac{1}{8^3}\frac{\pi^{5/2}}{6}\frac{c_\mathrm{s}^3}{G^{3/2}\rho^{1/2}},
    \end{equation}
    where $c_\mathrm{s}$ is the speed of sound of the gas, $\rho$ is the mass density and $G$ is the gravitational constant. At the sink-formation threshold, for a typical gas temperature of 200\,K the Jeans mass is $M_\mathrm{Jeans}\approx2000\,\Ms$ leading to a maximum cell mass of $M_\mathrm{c, Jeans}\approx4\,\Ms$.
    \item In order to accurately capture the expansion of the H\textsc{ii} regions around massive stars, it is necessary to resolve the 
    initial Str\"omgren sphere, which then proceeds to expand as a D-type ionization front. The size of this region is given by the Str\"omgren radius $R_\mathrm{Str}$ \citep{Stromgren}. We have chose to resolve the Str\"omgren sphere with at least 1000 cells, corresponding to approximately six cells per Str\"omgren radius. This consideration leads to a resolution criterion which we can write as
    \begin{equation}
      M_\mathrm{c,Str} = \frac{1}{1000}\frac{4}{3}\pi\rho R_\mathrm{Str}^3,
    \end{equation}
    which is equivalent to
    \begin{equation}
     M_\mathrm{c,Str} =4\times 10^{-4}\, \Ms \left(\frac{n}{10^4\,\dens}\right)^{-1} \left(\frac{\dot{N}_\mathrm{ion}}{10^{48}\,\mathrm{s}^{-1}}\right), 
    \end{equation}
    where $\dot{N}_\mathrm{ion}$ is the ionizing photon emission rate and $n$ is the nucleon number density. 
    \item Additionally we limit the maximum cell size within 10\,pc of the nearest sink particle to be no more than 0.15\,pc in diameter. This helps to better capture the behavior of low-density gas around the sink particles and ensures that the injection region for the SN energy is smaller than the Sedov-Taylor radius, which is
    \begin{equation}
        R_\mathrm{SD} = 24\,\mathrm{pc}\, n^{-0.42}E_{51}^{0.29},
    \end{equation}
    where $E_{51}$ is the explosion energy in units of $10^{51}\,\mathrm{erg}$. At our star formation threshold, i.e., at number densities of $n=10^4\,\dens$, the Sedov-Taylor radius is $R_\mathrm{SD}\approx 2\,\mathrm{pc}$.
    \item The size of the cells is limited at a minimum comoving volume of $0.1\,\mathrm{pc}^3\,h^{-3}$.
    \end{enumerate}
\begin{figure}
    \includegraphics[width=\linewidth]{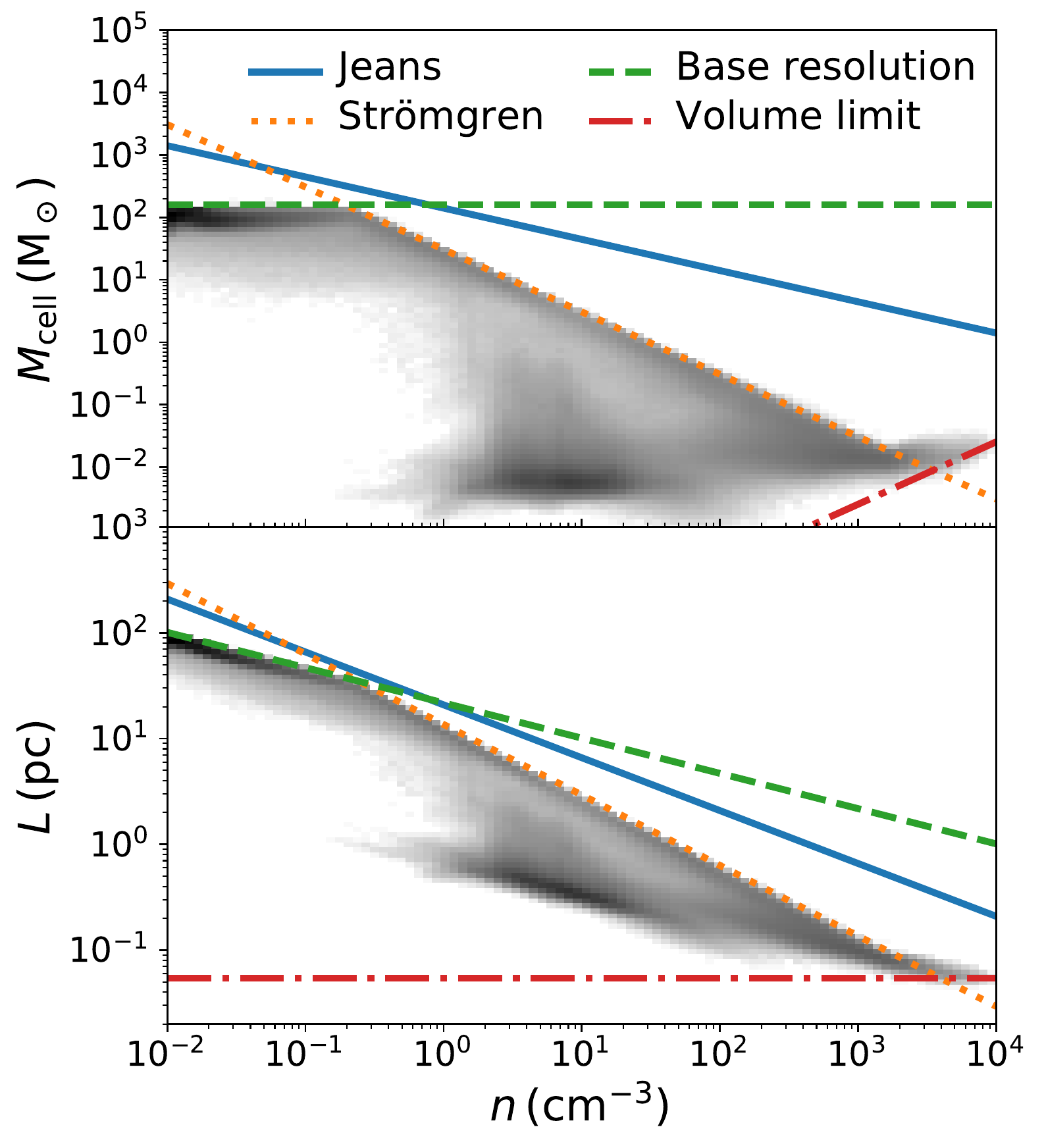}
    \caption{Comparison between the resolution criteria to resolve fragmentation (Jeans criterion, blue, solid) and to resolve photoionization (Str\"omgren criterion, orange, dotted), the initial gas particle mass (dashed, green), and the volume limit (dashed-dotted, red) as function of the nucleon number density. In the background, we show the distribution of cell sizes and masses in our simulation, when the first star forms. The resolution is expressed both as mass (upper panel) and as linear size (lower panel). }
    \label{fig:res}
\end{figure}
The refinement criteria dictated by the Jeans length and the Str\"omgren radius as a function of number density are shown in Fig.\ref{fig:res}. For this example we use an ionizing photon emission rate of $\dot{N}_\mathrm{ion}=4.0\times 10^{49}\,\mathrm{s}^{-1}$, which is appropriate for the 200\,\Ms\ stars that we model, and a gas temperature of $T=10000\,\mathrm{K}$ in the H\textsc{ii} region. It is worth noting that, despite the large ionizing radiation output, at number densities above $n=0.1\,\dens$\ the Str\"omgren criterion is more strict that the Jeans criterion. As the Str\"omgren criterion scales more steeply with density than the Jeans criterion, this will always be the case at sufficiently high densities. An unresolved ionized region leads to an over-cooling problem, similar to the over-cooling problem in unresolved SNe. Because the ionized region is not resolved, instead of a small fully ionized region we find a larger only slightly ionized region. The temperature and the thermal pressure of this region is underestimated. This slows or completely stops the expansion of the H\textsc{ii} region.

\section{Results} \label{sec:results}
\subsection{Qualitative description}
\begin{figure*}
    \includegraphics[width=\textwidth]{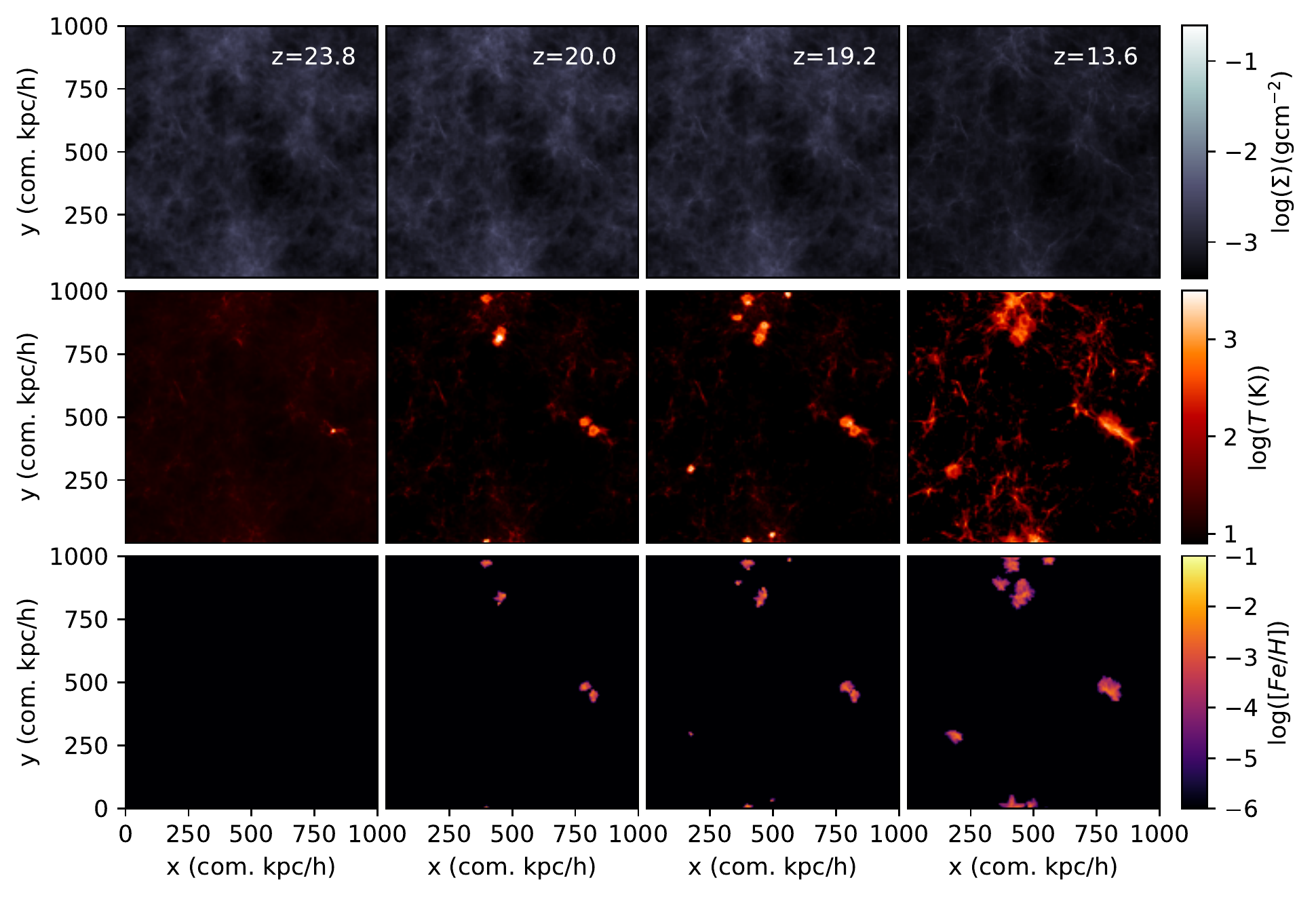}
    \caption{\label{fig:full_box}Column density (top row), temperature (middle row), and metallicity (bottom row) projections of the entire simulation at four different redshifts. The metallicities and temperatures are mass weighted averages along the column. Note that the third snapshot in this figure is the last time at which the radiative transfer is fully active (see text). A visualization of the simulation is included with online version of this study.}
\end{figure*}

\begin{figure*}
    \includegraphics[width=\textwidth]{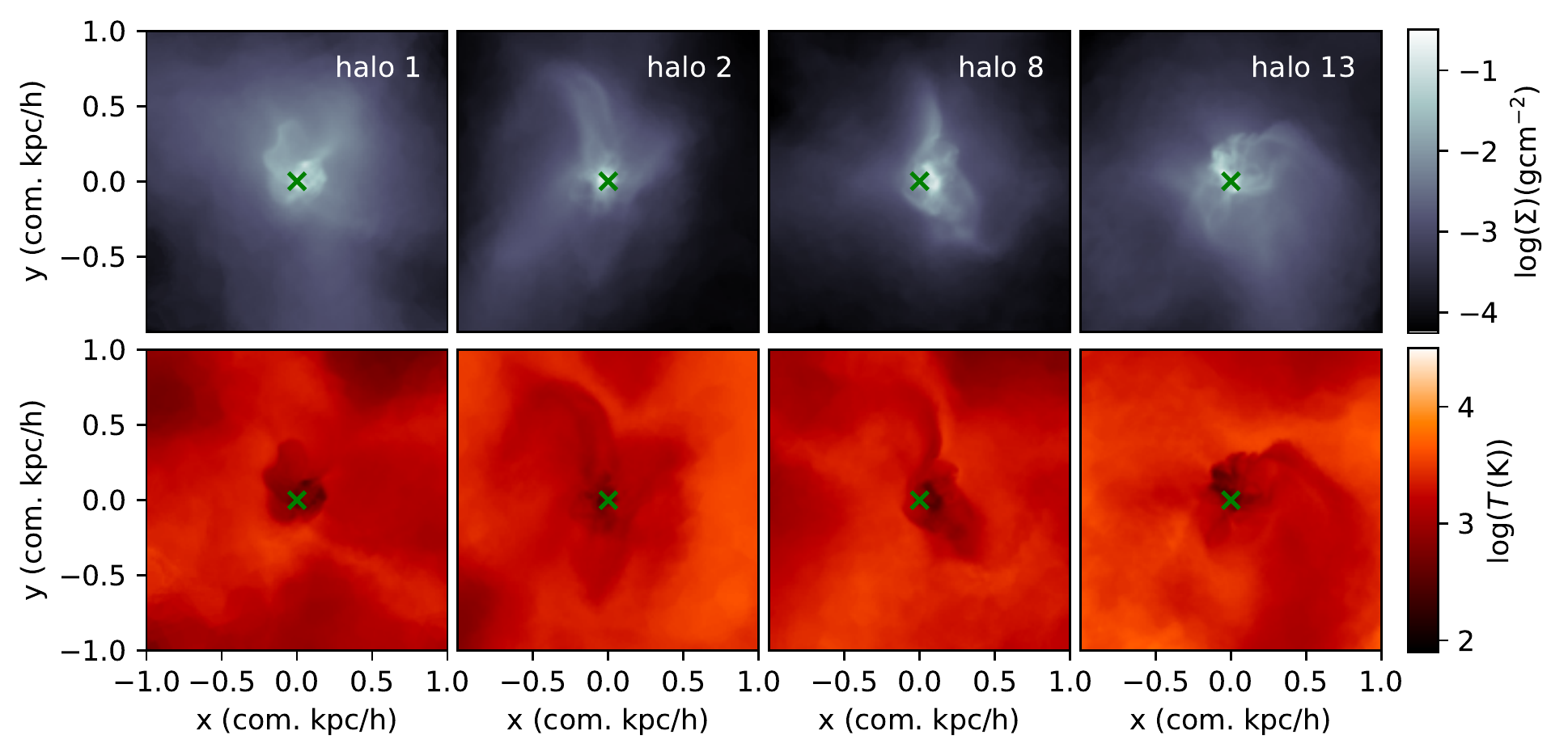}
    \caption{\label{fig:SF}Column density (upper row) and mass-weighted average temperature (lower row) at the onset of star formation in 4 selected halos. The individual projections are projections of a 2 kpc$\,h^{-1}$ cubes, centered on the sink particles, which are indicated with green crosses. Similar images for the remaining halos are found in Figures \ref{fig:apx_sf1} and \ref{fig:apx_sf2}. The halos we selected are two halos with short (halos 8 and 13) and two halos with long (halos 1 and 2) recollapse times. The properties of all halos are summarized in Table \ref{tab:halos}.}
\end{figure*}

Before the first metal-enriched star appears, we see Pop~III stars forming in 14 different halos, distributed throughout the simulation volume starting at $z\approx24$. We number these halos in order of appearance and summarize their property in Table \ref{tab:halos}. We show projections of the entire simulation box at various stages in Figure \ref{fig:full_box} and the centers of four selected halos in Fig. \ref{fig:SF}, two with long (halos 1 and 2) and two short (halos 8 and 13) recollapse times. Images of the remaining halos can be found in Appendix \ref{apx:more_details}. The radiation from the massive stars quickly photoevaporates the gas in their environment and drastically reduces the ambient density before they explode, similar to the behavior seen in previous simulations of radiative feedback from Pop~III stars \citep[see e.g.][]{wan04,WiseAbel2008}. While we allow multiple sinks, and hence multiple massive stars, to form in each halo, in practice the radiative feedback is so strong that this does not occur.

After the explosions we observe the formation of second-generation stars in three different modes: fast recollapse, slow recollapse and external enrichment. External enrichment implies that metal-enriched stars form in a halo that has never experienced metal-free star formation, i.e., that the metals come from outside the star-forming halo. Recollapse or internal enrichment refers to metal-enriched stars forming in halos that previously experienced metal-free star formation. Later, we will see that this appears to occur in two distinctly different modes, either relatively quickly after the SNe or after an extended time-period. We refer to these modes as fast and slow recollapse.

Only one of our halos (halo 4, seen in Figures \ref{fig:apx_sf1} and \ref{fig:apx_sn1}) does not recollapse during the 135\,Myr we simulate after the last SN. We observe 9 cases of internal and 8 cases of external enrichment. There is a discrepancy between the number of recollapses and the number of internal enrichments because four of the simulated halos merge with an externally enriched halo before forming the first internally enriched stars. These halos are therefore counted as recollapsed but, as the stars they form would be affected by the accreted second-generation stars, there is no formation of second generation stars in situ in these halos. The median recollapse time, i.e., the time between the first SN and the formation of second generation stars within the halo, is 50 Myr. 

\subsection{Evolution of halo masses}
As halo finders usually struggle with the high resolution of the gas in the centers of our halos, we determine halo masses only on the basis of their dark matter content. We use \textsc{rockstar} \citep{rockstar} to find halos in all snapshots and multiply the resulting halo masses with $\Omega_m/(\Omega_m - \Omega_b)$ to account for the baryons we excluded from the determination of the mass. Results from \citet{Schauer2019b} indicate that, given our streaming velocity, the baryonic gas mass can be lower than our estimate by a factor of two. However, our aim is to trace the underlying growth of the halos to find a connection between that growth history and the progression of early star formation, rather than, e.g., to quantify the effects of feedback on the baryon content of the halos. Therefore, and since baryons are only a secondary contribution to the overall mass of the halos, our assumption is appropriate for this analysis. 

\begin{figure}
    \includegraphics[width=\linewidth]{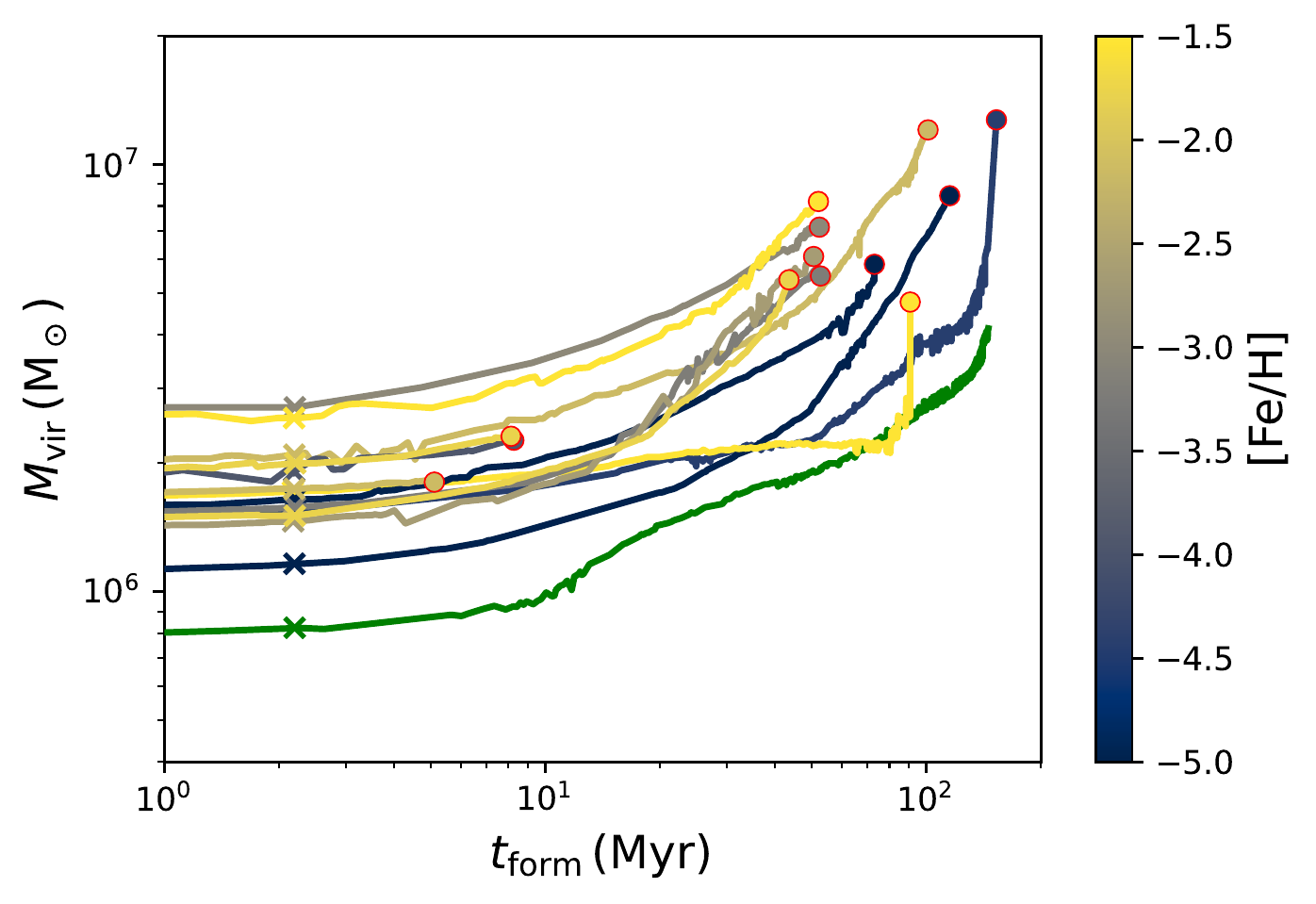}
    \caption{ \label{fig:M_vir} Growth history of the 14 halos that host Pop~III stars as a function of time since the formation of the sink particle. The halos are color-coded by the metallicity of the second generation stars that form in them. The green line is the halo that does not recollapse (halo 4). The crosses shows the halos shortly before the SNe and the circles show the time at which the first metal-enriched star forms in each halo.}
\end{figure}

We show the growth history of the 14 host halos until their recollapse in Fig. \ref{fig:M_vir}. The fitting function for critical halo-mass given by \citet[][their Eq. 9 \& 10]{Schauer2021} indicates that for our setup, i.e., streaming velocities of $v_\mathrm{bc} = 0.8\,\sigma$ and no global LW background, halos should start forming stars at $M_\mathrm{crit} \approx 2 \times 10^6\,\Ms$. We find a mean virial mass at star-formation of $M_\mathrm{mean, vir} = 1.7\times 10^6\,\Ms$ which is consistent with this previous results. As the fitting formula from \citet{Schauer2021} was obtained with higher resolution, this consistency shows that our resolution is sufficient to capture the collapse of the minihalos without artificial delays caused by low resolution.

There is no simple relationship between halo mass and recollapse time or metallicity, indicating that mixing and recollapse are stochastic processes. We will discuss the metallicities in more detail in Section \ref{sec:met}. The halo that does not recollapse seems to have a relatively low mass, and exhibits slow growth. Three of the halos recollapse in less than 10\,Myr, yet this behavior does not correspond to the highest halo masses or particularly fast halo growth. The halos that recollapse after a longer time usually at least triple in mass before falling back. The least massive halo does not recollapse during our simulation. The complexity of this picture demonstrates the need for simulating these processes in a statistical sample, rather than only simulating one example of a halo of a certain mass.

\subsection{Pre-SN environment}
\label{sec:pre_sn}
\begin{figure*}
    \includegraphics[width=\textwidth]{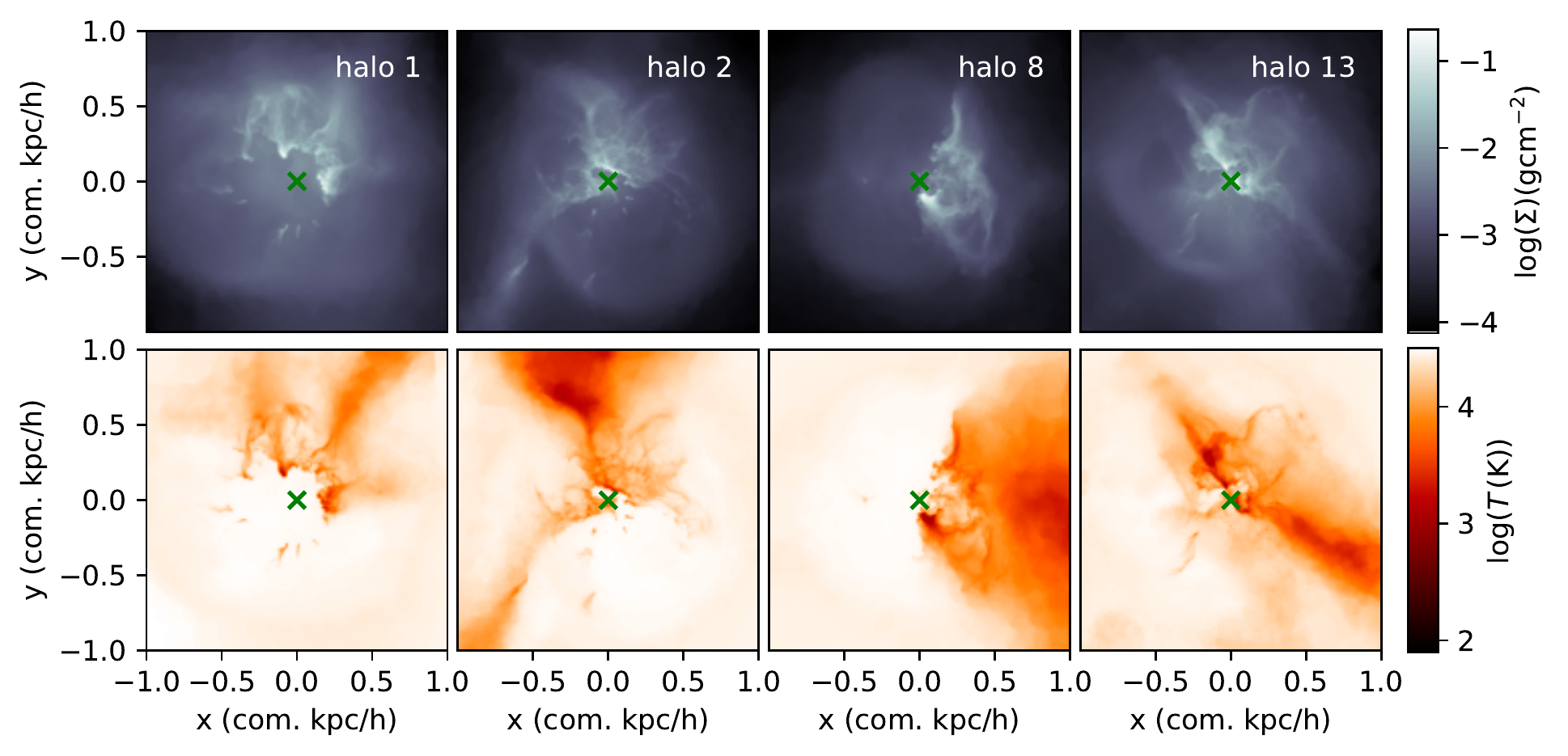}
    \caption{\label{fig:pre_sn_imgs}Column density (upper row) and mass-weighted average temperature (lower row) shortly before the SNe in 4 selected halos shown in Figure \ref{fig:SF}. The green crosses indicate the position of the Pop~III stars that are about to explode. We can see that, while being mostly ionized, the halos retain small, dense clumps and shadowed cool regions behind them. Similar images for the remaining halos can be found in Figures \ref{fig:apx_sn1} and \ref{fig:apx_sn2}.}
\end{figure*}
We show the four selected halos shortly before their SNe in Fig \ref{fig:pre_sn_imgs}. Halo 1 has the least homogeneous mixing among the internally enriched halos. This can be seen in Table \ref{tab:halos}, as this halo has the largest difference between local metallicity in the star forming region and the total metallicity of the halo. We can see that the halos are mostly photoionized, hot and photoevaporated. However, there are localized dense clumps that are relatively cold, and cool, shadowed regions behind these clumps. We note that not all halos exhibit an equal degree of clumpiness. Halo 4, which does not recollapse during our simulation, has only very little dense gas left before the SN (see Fig. \ref{fig:apx_sn1}).

\begin{figure}
    \includegraphics[width=\linewidth]{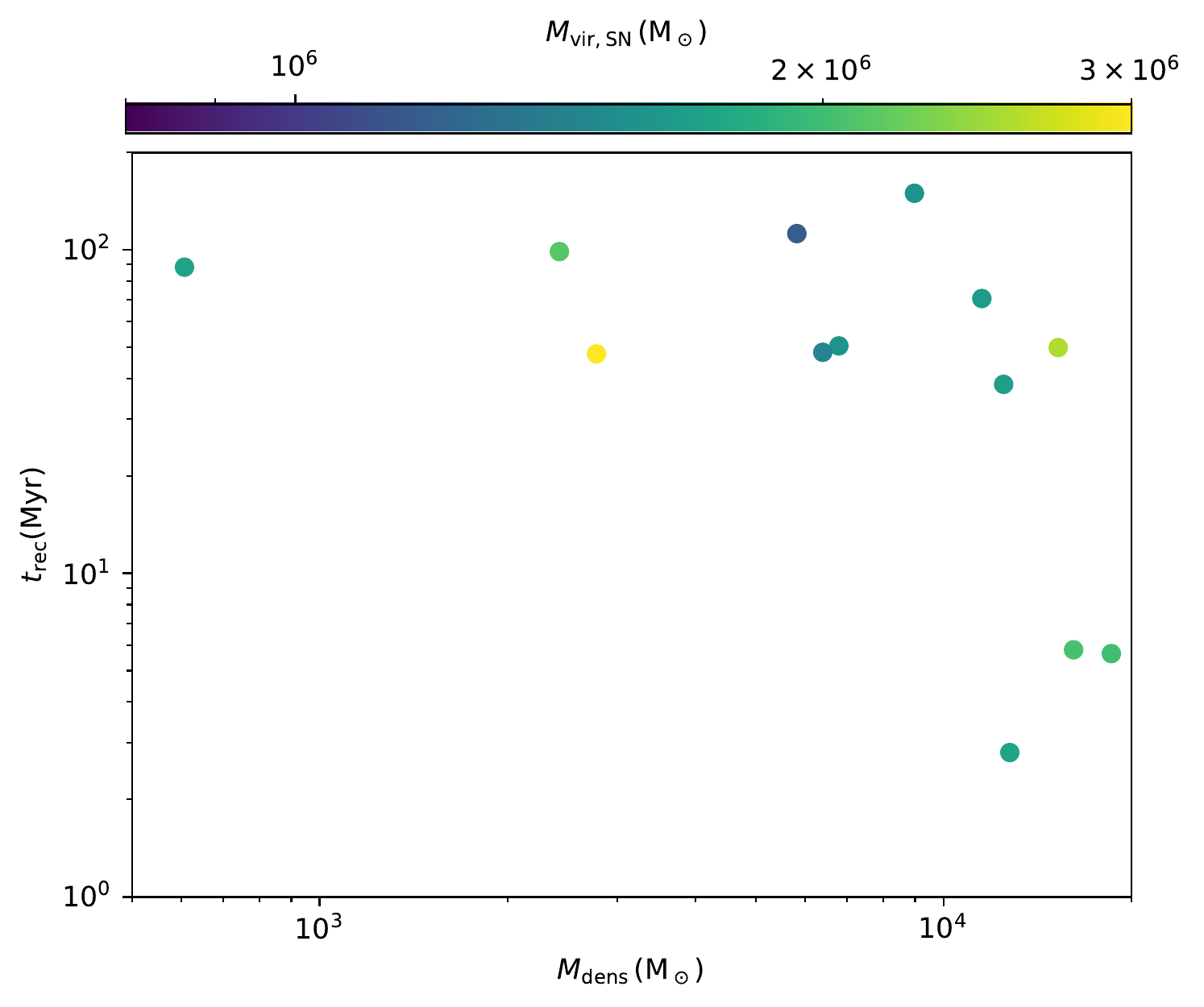}
    \caption{Fall-back time as function of the dense mass, i.e., the mass at nucleon number densities above 1000\,\dens\ within the halo. The dense masses are measured shortly before the SNe. The colors correspond to the virial masses of the host halos.} 
\label{fig:t_fall}
\end{figure}

In Fig. \ref{fig:t_fall}, we examine whether there is a correlation between the presence of clumps shortly before the SN and the recollapse time. For this purpose we define dense gas as having a nucleon number densities above 1000\,\dens. Short recollapse times are always associated with large amounts of dense gas. However, the contrary is not necessarily true, not all such with large dense gas masses undergo star formation immediately. Some of them are disrupted by the SNe, leading to long recollapse times of the order of 100\,Myr. Exploring what properties of the halos or the clumps ultimately determine the recollapse time exceeds the scope of this work and will be subject to a future study.

Halos without high-density clumps always have long recollapse times. There is no obvious correlation with the mass of the halos, as both very high and very low-mass halos are found among the ones with long recollapse times and the halos with short recollapse times are in an intermediate mass range. Therefore, the baryonic sub-structure of halos has a stronger influence on recollapse time than the mass of the halo. 

\subsection{Metallicities}
\label{sec:met}
\begin{figure*}
    \includegraphics[width=\textwidth]{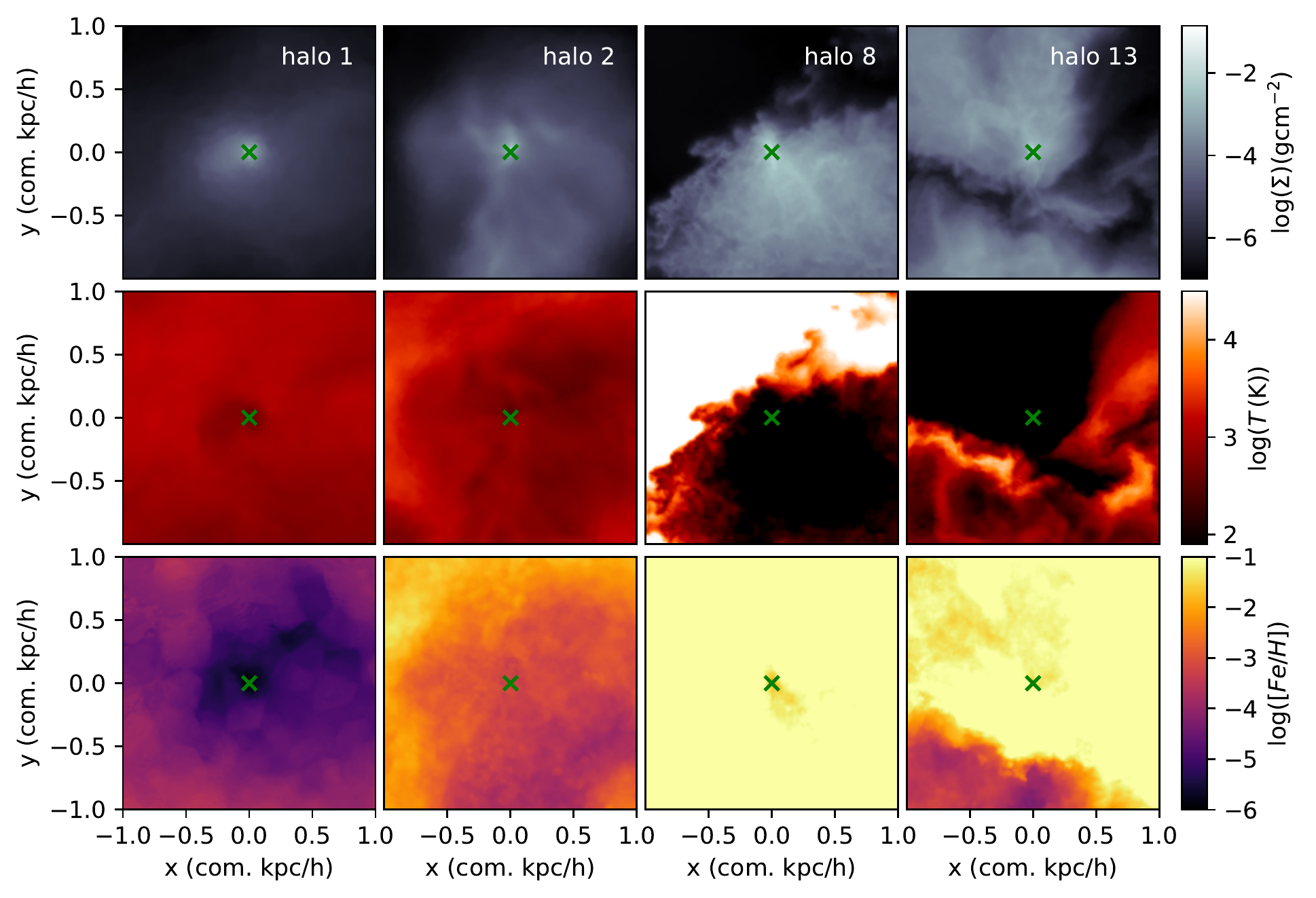}
    \caption{\label{fig:rec}Column density (upper row), average temperature (middle row) and metallicity (bottom row) at recollapse for four internally enriched halos. The second generation stars are indicated with green crosses. We can see a variety of different mixing behaviors. Some halos (e.g., halo 8) have a relatively homogenous metallicity, whereas others (e.g., halo 1) show a steep decrease in metallicity towards the center of collapse. Similar images for the remaining halos can be found in Fig. \ref{fig:apx_rec}.}
\end{figure*}

\begin{figure*}
    \includegraphics[width=\textwidth]{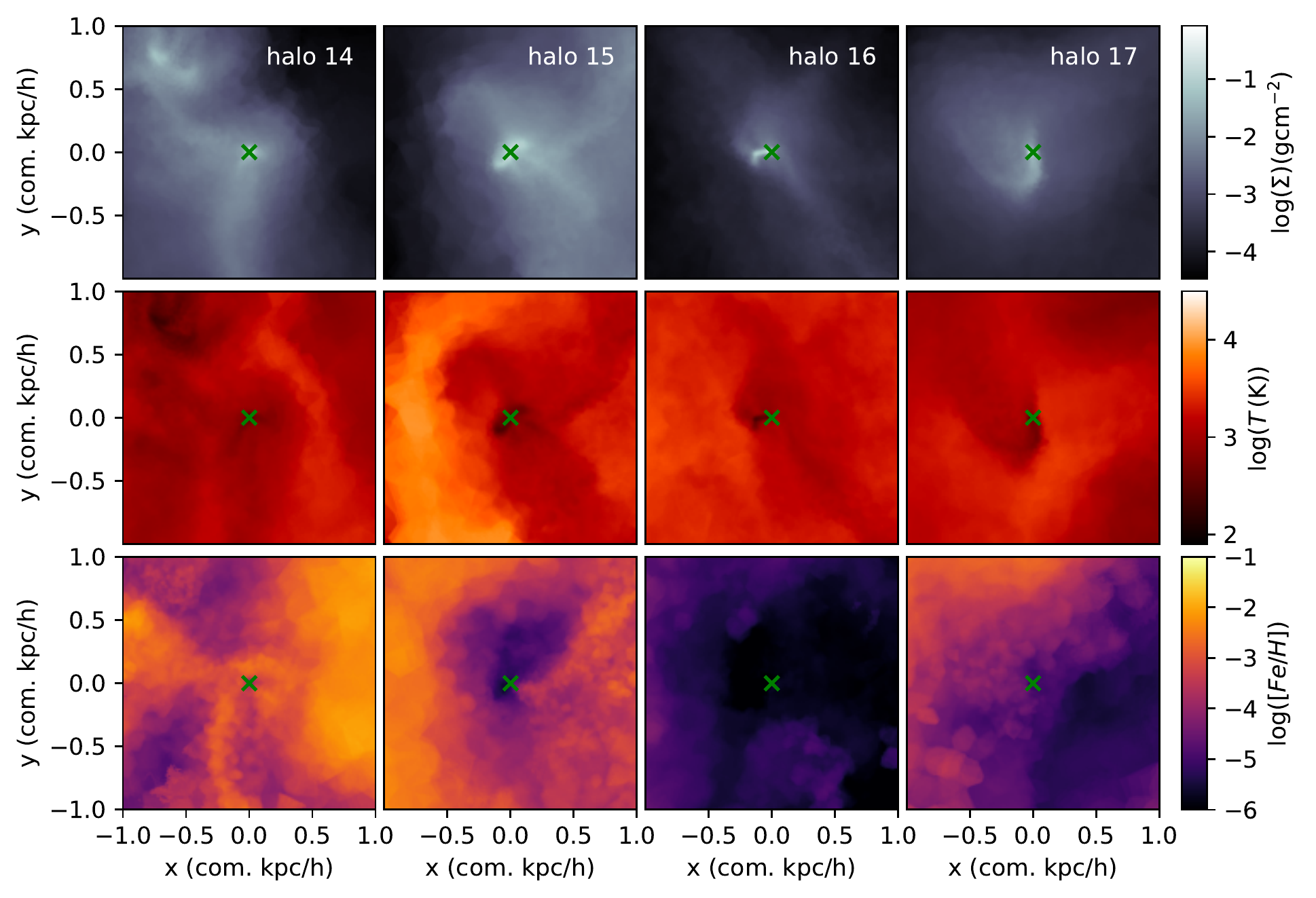}
    \caption{\label{fig:ext}. Same as Fig. \ref{fig:rec} but for four of the externally enriched halos. Images of the remaining halos are shown in Fig. \ref{fig:apx_ext}.
    } 
\end{figure*}

Metal-enriched stars form in both internally and externally enriched halos. Projections of internally enriched halos at the moment of second-generation star formation are shown in Fig. \ref{fig:rec}, and externally enriched halos in Fig. \ref{fig:ext}. The figures demonstrate that second-generation star formation occurs in a variety of different environments. In some cases (e.g., halo 1) the environment is cold, and the star forms in a clump that is less enriched by metals than the surroundings. Other cases (e.g., halo 8) show star formation in a high-metallicity environment next to an H\textsc{II} region. The halo-masses of these externally enriched halos can be found in Table \ref{tab:ext_halos}.

We determine the metallicities of the second generation stars that form in the simulation by counting the number of tracer particles ($N_\mathrm{tr}$) within the sink-formation radius around each newly-formed sink particle and comparing this with the total baryonic mass enclosed within the same volume ($m_\mathrm{b}$). The metallicity follows from these quantities as
 \begin{equation}
[\mathrm{Fe}/\mathrm{H}] = \log_{10}\left( \frac{N_\mathrm{tr}\times m_\mathrm{Fe, SN}}{N_\mathrm{tr,SN} \times m_\mathrm{b}}\right) - \log_{10}(f_\mathrm{Fe,\odot}),
\end{equation}
where $N_\mathrm{tr,SN}=10^7$ is the total number of tracer particles injected into each SN, $m_\mathrm{Fe,SN} = 56\,\Ms$ is the mass of iron produced by the supernova \citep{HegerWoosley2002}, and $f_\mathrm{Fe,\odot} = 0.00134$ is the solar iron abundance \citep{Asplund09}. This assumes that the initial SN ejecta are well-mixed, i.e., that all elements are distributed evenly throughout the ejecta. It is unclear whether this assumption accurately describes the mixing behavior in real SN ejecta, but given the enormous difficulties involved in modelling the resulting metal distribution in 3D supernova explosions, it remains the most conservative assumption.

Our main result is that PISNe can lead to the formation of second-generation stars with a broad range of metallicities. Some indeed exhibit high metallicities in the range $-2<[\mathrm{Fe}/\mathrm{H}]<-1$, as predicted by the models of \citet{Karlsson2008}, \citet{deBennassuti17} and \citet{Salvadori2019}. However, most second generation stars in our simulation would be classified as EMP stars, with metallicities of $[\mathrm{Fe}/\mathrm{H}]<-3$. 

We find that externally enriched stars always have low metallicities ($[\mathrm{Fe}/\mathrm{H}]<-3$) in our simulation, whereas internally enriched stars have the highest metallicities, reaching $[\mathrm{Fe}/\mathrm{H}]>-2$. Five out of nine internally enriched stars have metallicities higher than the EMP threshold, three stars exhibit metallicities just below that range with $-4<[\mathrm{Fe}/\mathrm{H}]<-3$, and one star, with the lowest internal enrichment of all, falls in the range of $[\mathrm{Fe}/\mathrm{H}]<-6$.

There are two obvious possible reasons for the low metallicities we find: either most of the metals are ejected from the halos or the metals are not fully mixed into the star-forming gas. In the latter case we would expect the metallicity of the second generation stars to be lower than the average metallicity of the halos. This is referred to by \citet{Chiaki18} as inefficient internal enrichment.

\begin{figure}
    \includegraphics[width=\linewidth]{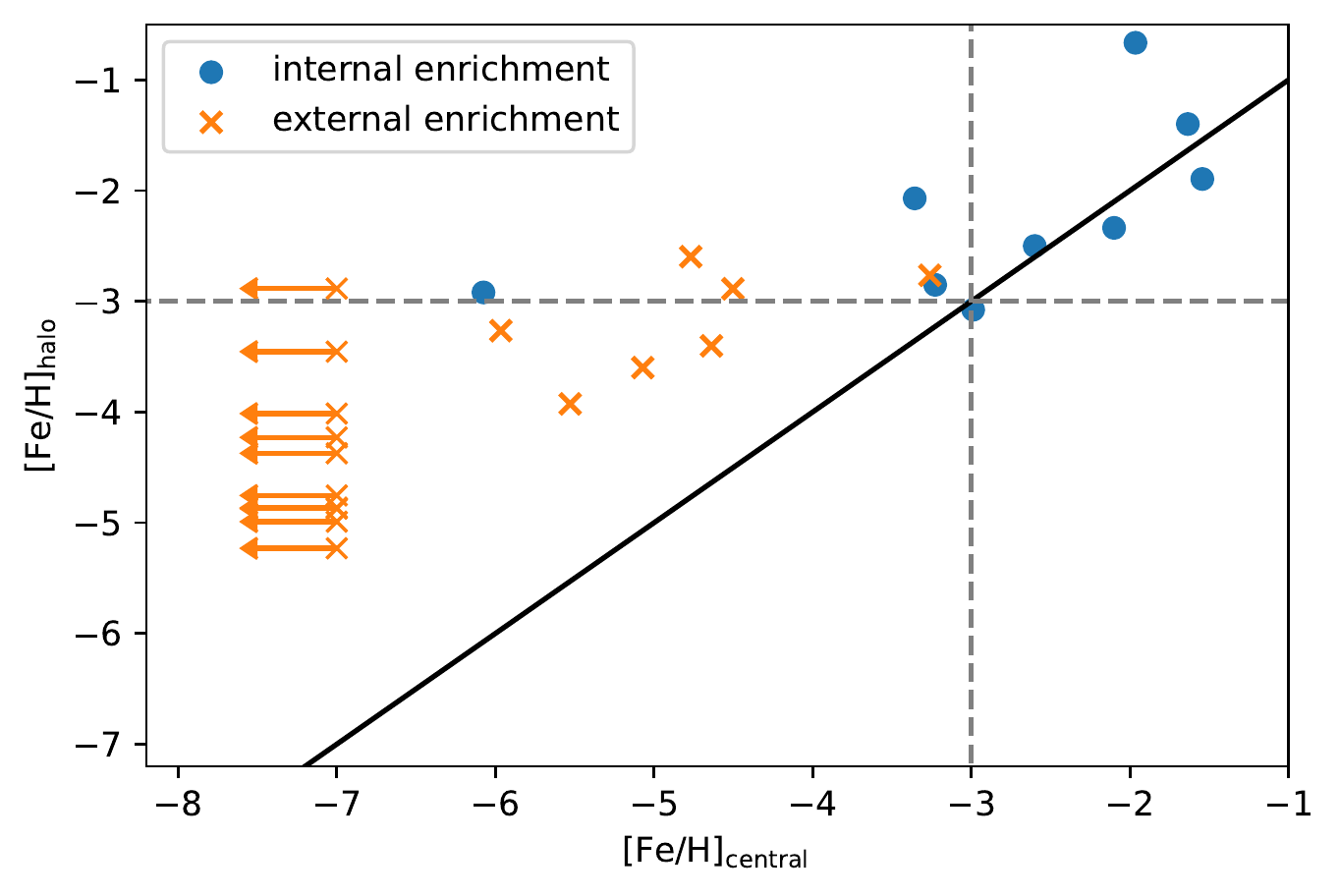}
    \caption{Metallicity of second generation stars compared to the metallicity of their host halos. Crosses indicate external enrichment and dots indicate internal enrichment. The upper limits set at $[\mathrm{Fe}/\mathrm{H}]=-7$ represent cases in which the halo is only superficially enriched, yet the central star forming region does not contain metals. The grey dashed line indicates the limit for what is considered to be EMP stars, i.e., [Fe/H]$=-3$.}
    \label{fig:Z_halo}
\end{figure}

\begin{figure}
    \includegraphics[width=\linewidth]{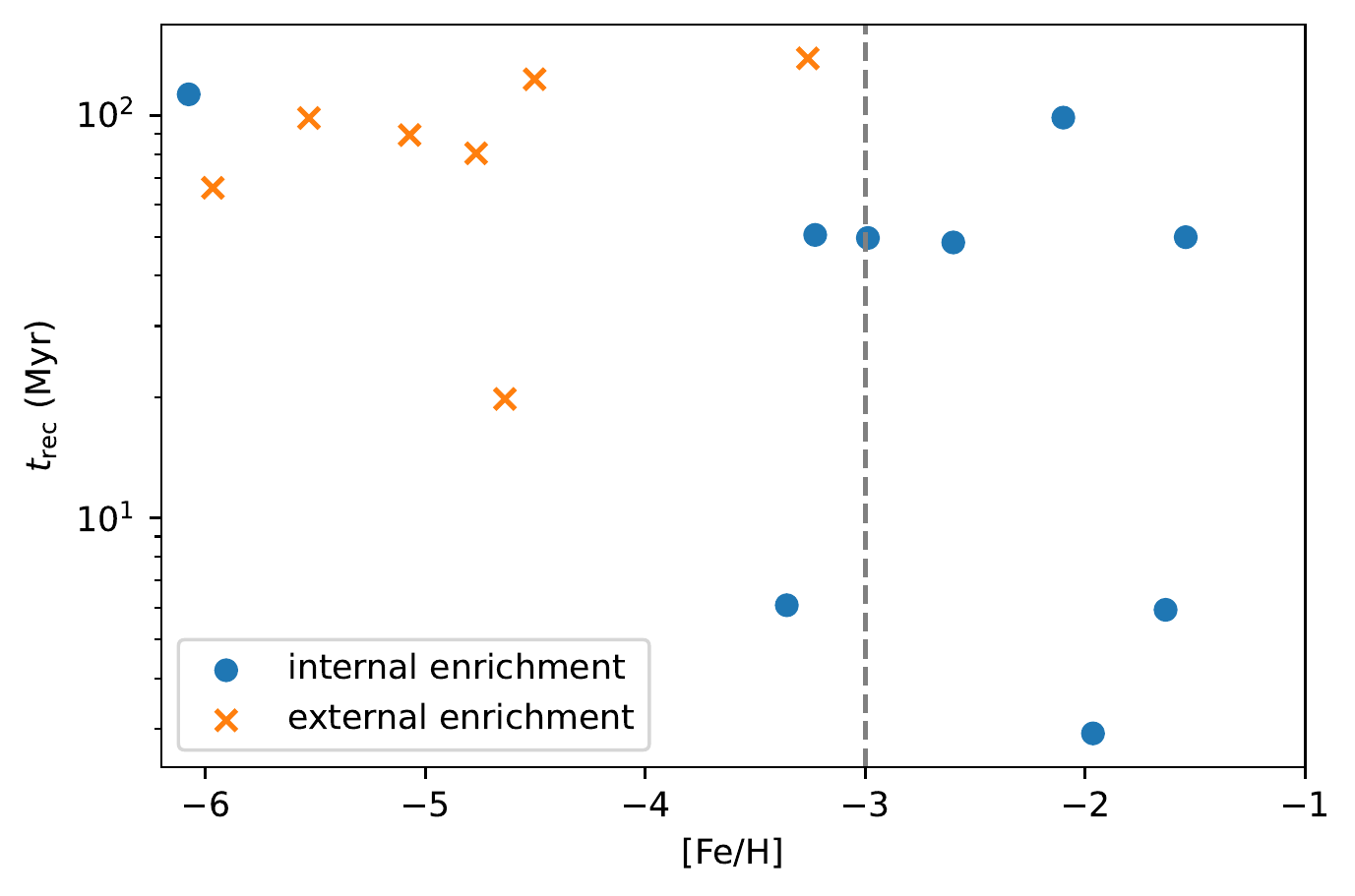}
    \caption{Metallicity of second generation stars as function of time since the SN. The crosses indicate external enrichment, the dots indicate internal enrichment and the grey dashed line represents the limit of what constitutes an extremely metal-poor star.}
    \label{fig:t_Z}
\end{figure}

In order to determine which of these two reasons is mainly responsible for the low metallicities, we compare in Fig. \ref{fig:Z_halo} the average halo metallicities to the metallicities of the second generation stars forming in the center of the halos. We find that in most cases of internal enrichment the halos are well mixed. The ejection of metals therefore is dominant in these halos. There are two cases which show lower central than halo-scale metallicity: one is halo 1 at $\FeH \approx -6$ and the second one is halo 12 at $\FeH\approx -3.4$. However, as can be seen in Fig. \ref{fig:t_Z} only the $\FeH \approx -3.4$ case corresponds to a short recollapse time and therefore indicates inefficient internal enrichment. The $\FeH \approx -6$ case has a recollapse time of more than 100\,Myr. This star cannot have formed from a clump in the host halo which survived the SN, as such a clump would have a much shorter free-fall time. Thus, it is more likely that this is actually a case in which an externally enriched halo merges with the original host shortly before star formation, such as reported by \citet{bsmith15}. We will investigate this particular star and what conditions lead to its formation in a future study.

In the externally enriched halos, the central metallicity is commonly much lower than the average metallicity of the halos. The are many cases in which the halos are only superficially enriched but their star-forming centers remain pristine. 

\section{Discussion} \label{sec:discussion}
\subsection{Streaming Velocities}
In this paper, we include streaming velocities in a simulation that investigates the spread of metals from Pop~III stars for the first time. A streaming velocity of 0.8\,$\sigma_{\rm rms}$ is the most common value in the Universe; regions without a streaming velocity are very rare. We confirmed that external enrichment does not preferentially occur in the direction of the streaming velocity, implying there is no significant effect on the transport of metals through the intergalactic medium. Thus, any effect of the streaming has on external enrichment acts via the properties of the haloes, such as the change in the size, the structure and the baryonic mass content. It is known that streaming velocities increase the minimum halo mass for Pop~III star formation \citep{Schauer2021}, but decrease the gas content in these halos \citep{Naoz13}, and it is unclear how much a larger potential well decreases the escape of metals for internal enrichment. On the other hand, the shape of the gas component of minihalos changes: the gas is more oblate and less spherical in streaming velocity regions compared to non-streaming velocity regions \citep{Druschke20}. In the three-dimensional halo structure, this might lead to more escape paths for ionizing radiation and subsequently the spread of metals into the surrounding. We speculate that external enrichment could be enhanced in streaming velocity regions but testing and investigating this issue is a task for future work. 
\subsection{Mixing, enrichment and dilution}
Our simulation shows a variety of different behaviors with regards to metal enrichment, mixing, dilution and the formation of second generation stars. A key question here is how much the SN ejecta are mixed with the pristine ambient medium before second-generation stars can form. There have been numerous studies attempting to investigate this issue with analytical and numerical models. We compare our results to these earlier studies in this section.

\begin{figure}
    \includegraphics[width=\linewidth]{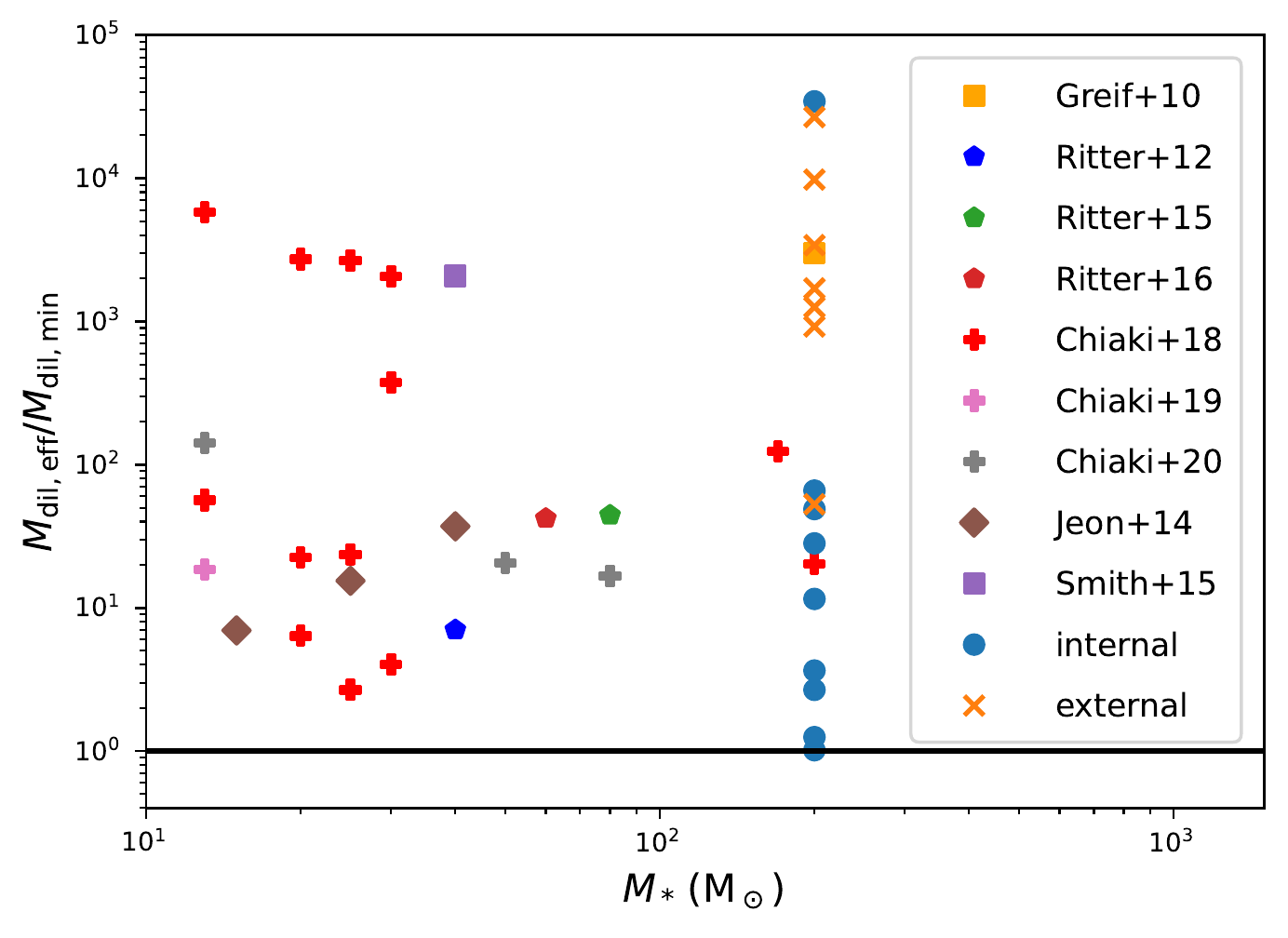}
    \caption{Comparison of the metal dilution found in our simulations and in the literature with the analytic lower limit derived by \citet{Magg20}. We show the ratio of the effective dilution mass to the analytic limit, as function of the stellar mass of the supernova progenitor. The analytic limit for our PISNe is $M_\mathrm{dil, min} = 1.6 \times 10^6\,\Ms$. Results from this work are shown as blue circles (internal enrichment) and orange x-symbols (external enrichment). The other symbols show values from a range of previous studies, as indicated in the key. In every case, the ratio is greater than 1, indicating that the results are consistent with the analytic limit.}
    \label{fig:m_dil}
\end{figure}

\citet{Magg20} recently derived a lower limit for the mass of pristine material that the ejecta from the first SNe are diluted into. To compare our results to this limit, we compute effective dilution masses, i.e., the mass of pristine material the SN ejecta would have to mix with homogeneously in order to arrive at the found metallicity for all of our second generation stars. In Fig.~\ref{fig:m_dil}, the dilution masses are compared to the analytical lower limit of 
\begin{equation}
    M_\mathrm{dil, min} = 1.9\times 10^4\,\Ms\, E_{51}^{0.96}= 1.6 \times 10^6\,\Ms,
\end{equation}
as derived in \citet{Magg20}. Here, $E_{51}$ is the explosion energy of the SN in units of foe. The Figure is an adaptation of figure 1 in \citet{Magg20} and contains values from the simulations by \citet{Greif10, Ritter12, Ritter15, Ritter16, Jeon14, Chiaki18, Chiaki19, Chiaki2020} and \citet{bsmith15} for comparison. Despite our dilution masses scattering over five orders of magnitude, our lowest value is 2\% above the analytical limit, and therefore our simulation is consistent with it.

The model by \citet{Chiaki16} predicts that our halos are at the border between internal and external enrichment, and thus we would expect to see both in our simulation. This expectation is consistent with our findings. However, according to this model, for a 200\,\Ms\ Pop~III star, inefficient internal enrichment and the associated short recollapse times should only be possible in halos more massive than $M_\mathrm{vir} = 2\times 10^7\,\Ms$. From the fact that we see both, inhomogeneous mixing as well as short recollapse times, in halos around one order of magnitude less massive than this threshold we conclude that inefficient internal enrichment is governed by the presence of complex baryonic sub-structures within the halo rather than by the total halo mass.

That recollapse occurs quite frequently stands in contrast to the picture painted by \citet{Whalen08b} in which halos do not recollapse. This difference is likely caused by two factors. Firstly, as the simulations by \citet{Whalen08b} are one-dimensional the ionizing radiation and the thermal SN energy can only get away from the star by evaporating the halo. Whether a halo becomes fully or only partially evaporated depends solely on the balance between the binding energy and the energy injected by stellar feedback. In our model a large fraction of the halo gets evaporated quickly, allowing radiation and injected thermal energy to escape before all of the dense interior of the halo has been destroyed. This significantly weakens the efficacy of stellar feedback in preventing future star formation. Secondly, the halos in \citet{Whalen08b} lack the cosmological context. While our halos can accrete new gas via mergers or along the cosmic filaments, the one-dimensional halos can only re-accrete gas that has been heated by the SN. \citet{Chiaki16} find that this mechanism is an important factor in controlling the recollapse. 

Semi-analytical and semi-numerical models of metal-mixing after Pop~III formation often assume homogeneous mixing or a parametrized mixing model \citep{Karlsson2008, deBennassuti17, Hartwig18b, Salvadori2019}. Due to the large yields of PISNe, second generation stars are often relatively metal-rich ([Fe/H]$\sim-2$) in these models. While we see instances of this behavior, we typically find much lower metallicities. This decrease is due to large inhomogeneities in mixing, especially in externally enriched halos, due to mixing with a large amount of pristine material and due to the ejecta of metals from the halos.

Generally, we find that internally enriched halos have a more homogeneous metallicity distribution than externally enriched halos. \citet{Tarumi2020} developed a model for this effect and find that also in larger simulations \citep[by][]{Xu16}, externally enriched halos have lower metallicities in their star-forming centers. The reason for this is that metals do not easily mix into dense centers of already formed halos from the outside \citep{Chen17a}.

\subsection{Where are the stars with PISN patterns?}
If massive PISNe with a strongly pronounced odd-even effect are indeed common in the early Universe, our simulations would pose a conflict with the observed abundance patterns of EMP stars. Thus far, there is only one candidate star found that could be enriched by a PISN \citep{Aoki14}, and it is not in the EMP regime. We have seen that two of the primary hypotheses \-- those of large metallicities and the destruction of halos \-- cannot explain this discrepancy. Isolated massive PISN should lead to the formation of EMP stars, which would be commonly recovered in surveys targeting these stars \citep[such as][]{Youakim2020, TOPoS5}. 

There are several remaining explanations. The most obvious one is that sufficiently massive stars do not form or are exceedingly rare. Only a combination of forward modelling the formation of PISN-enriched stars in the progenitors of the Milky Way and a quantitative analysis of observed abundance patterns can constrain how rare exactly PISNe would have to be to account for the current lack of detections of PISN-enriched stars. 

Additionally, if massive Pop~III stars do not form in isolation but instead are close to other stars that explode as type II SNe, our predictions may not hold as well. There have so far been only very limited studies that investigate the mixing of ejecta from several Pop~III SNe \citep{Ritter15} and a set of dedicated detailed simulations with a sufficiently large sample of halos will be necessary to investigate this scenario. One difficulty to overcome in this scenario is that PISNe have very large yields and therefore, if mixed homogeneously, easily dominate the metal abundances in stars enriched by several SNe \citep{Hartwig18a}. However, we see such a rich diversity of mixing outcomes, that completely homogeneous mixing seems unlikely. \citet{Cooke2014} suggest, based on an analytical model, that if several SNe explode in the same halo the resulting abundance pattern will likely be more reflective of the less energetic SNe. This effect is caused by the more energetic material leaving the halo preferentially. A detailed numerical investigation of this scenario will be necessary to fully understand how the presence of multiple stars in the same halo affects the outcome.

For the second generation stars to survive until and be observable at the present day, they need to be low enough in mass. At which metallicity the transition from high-mass Pop~III-like star formation to a bottom-heavy more Salpeter-like IMF occurs is still an open question \citep{GreifReview}. Metal-line cooling starts to be an effective source of cooling around metallicities of $[Fe/H]\approx -3$. However, low-mass star formation must be possible at lower metallicities, since lower-metallicity stars have been observed \citep[see e.g.][]{Nordlander2019}. The additional source of cooling that may facilitate low-mass stars to form at these metallicities is dust cooling. However, at which metallicity dust cooling becomes efficient depends on the amounts, kinds, and grain-size-distribution of dust in the SN ejecta \citep{Schneider2012, Chiaki17}. Therefore, modelling dust-grain formation in PISN remnants will be a crucial step in understanding the implications of stellar archaeology for PISNe. The critical metallicity and general conditions required for low mass star formation are not yet well-understood. While observations of stars with extremely low metallicities show that it must be possible to form low-mass stars in the metallicity range seen in our model, it is still unknown how prevalent such low mass star formation is and what conditions it requires. Thus, the possibility that all stars carrying PISN fingerprints are short-lived can currently not be ruled out. 

Finally, we note that our conclusions apply only to non-rotating, massive, and energetic PISNe. Stars that are close to the PISN mass-limit, i.e., stars with masses around 150\,\Ms, are expected to have substantially smaller explosion energies and yields \citep{Gilmer2017}. Also stellar rotation can have a crucial impact on the evolution and explosion of metal-free stars \citep{Meynet_rev, Murphy2021}.

\subsection{Caveats and extensions}
Building up on the work presented here, our simulation could be extended to cover the formation of the first galaxies, rather than only considering the first two generations of stars. In order to do this, the simulations will need to determine the metallicity of a newly forming star at runtime, rather than in the current post-processing step. Improving this aspect would allow us to investigate mixing of metals from SNe of different populations of stars. We are limiting this simulation to the extreme case of 200\,\Ms\ Pop~III stars with PISNe. Future simulations will sample stars from a more realistic Pop~III IMF, and make predictions of second-generation stars arising from an specific IMF.

While we are modelling local radiative feedback with high physical fidelity in the LW and ionizing bands, our implementation of LW feedback is less advanced. Firstly, we do not include LW self-shielding. \citet{Kitayama2004} and \citet{Schauer15, Schauer17a} have shown that extremely high spectral and spatial resolution is required to accurate model the escape of LW radiation from minihalos. It is questionable whether such high resolutions can be achieved in three-dimensional cosmological simulations in the intermediate future. Secondly, we currently do not account for a global LW background. Choosing a global LW background and its evolution with redshift will require careful consideration, as it depends on escape fractions, the large-scale star formation rate and the IMF and the spectra of the forming stars. As LW radiation has a much longer mean free path than our simulated volume, modelling this background self-consistently is currently not possible.

\section{Summary} \label{sec:sum}
In this study we use a cosmological simulation to model the formation of the first stars, their feedback, their explosion as PISNe and afterwards the formation of the first metal-enriched stars. Our findings are:
\begin{enumerate}
    \item In the 1\,Mpc$\,h^{-1}$ simulation box, there are 14 halos forming one Pop~III star formation each before $z=19$. In accordance with previous predictions, our halos form stars at a virial mass of $M_\mathrm{vir}\approx 2\times 10^6\,\Ms$.
    \item At $z=19$ we deactivate stellar feedback and wait for the halos to recollapse. In the following 135 Myr most of our halos recollapse and we see various instances of externally and internally enriched formation of metal-poor stars. 
    \item Halos either recollapse quickly ($t_\mathrm{rec}< 10\,\mathrm{Myr}$, in three cases), slowly ($t_\mathrm{rec} > 40\,\mathrm{Myr}$, in nine cases) or not at all ($t_\mathrm{rec} > 135\,\mathrm{Myr}$, one case). Which of these paths they take depends on their baryonic substructure at the time of the SN, but it is not yet clear what exact criteria are required to enable the fast recollapse. 
    \item While the least-massive halo does not recollapse, among those that do recollapse, we find no direct correlation between the halo mass and the recollapse time.
    \item Second generation stars in externally-enriched halos start forming 70\,Myr after the SNe.
    \item We find a wide variety of metallicities of the second generation stars. Out of the sixteen second generation star-forming regions we find three have relatively high metallicities ($\FeH \geq -2$) and two have intermediate metallicities ($-2 < \FeH < -3$). All of these are formed by internal enrichment. Most of our second generation star-forming regions (eleven out of sixteen) are in the EMP ($\FeH \leq -3$) regime. Most of the lowest-metallicity stars are formed by external enrichment.
    \item The effective dilution mass scatters over 5 orders of magnitude. In all cases, our results are consistent with the analytical minimum of $1.6\times 10^6\,\Ms$ \citep{Magg20}.
\end{enumerate} 
Since PISNe are expected to produce a very characteristic pattern of elemental abundances, and since this pattern is rarely observed, these results stand in tension with the measured elemental abundances in extremely metal-poor stars. This may indicate that PISNe of the kind we investigate have been exceedingly rare in the early Universe. Possible alternative scenarios are mixing of several SNe, low mass stars not being able to form in the lower metallicity gas, or less characteristic SN yields. We will investigate the first of these alternatives in a follow-up study.

\begin{acknowledgements}
We would like to thank the anonymous referee for providing helpful and constructive feedback on this manuscript. The authors thank Volker Springel for sharing the \textsc{Arepo} hydrodynamics code with them. We also thank R\"udiger Pakmor and Paul Clark for help, technical support and advice. MM was supported by the Max-Planck-Society through the fellowship program of the International Max-Planck Research School for Astronomy in Heidelberg (IMPRS-HD). Support for this work was provided by NASA through the Hubble Fellowship grant HST-HF2-51418.001-A, awarded by STScI, which is operated by AURA, under contract NAS5-26555. RSK acknowledges support from the Heidelberg cluster of excellence EXC 2181 (Project-ID 390900948) ”STRUCTURES: A unifying approach to emergent phenomena in the physical world, mathematics, and complex data” funded by the German Excellence Strategy. RSK, SCOG, and RGT acknowledge financial support from the German Research Foundation (DFG) via the collaborative research centre (SFB 881, Project-ID 138713538) ”The Milky Way System” (subprojects A1, B1, B2, and B8). 

The simulations were performed on Supermuc-NG and the authors gratefully acknowledge the Gauss Centre for Supercomputing e.V. (\url{www.gauss-centre.eu}) for funding this project by providing computing time on the GCS Supercomputer SuperMUC-NG at Leibniz Supercomputing Centre (\url{www.lrz.de}) under computing grant number pr74nu. The authors also gratefully acknowledge the data storage service SDS\@hd and the computing service bwHPC, supported by the Ministry of Science, Research and the Arts Baden-Württemberg (MWK) and the German Research Foundation (DFG) through grant INST 35/1314-1 FUGG. 
\end{acknowledgements}

%

\vspace{5mm}
\facilities{Gauss:LRZ}


\software{Rockstar \citep{rockstar},
          NumPy \citep{Numpy},
          SciPy \citep{SciPy},
          h5py \citep{h5py},
          Arepo \citep{arepo}
          }



\appendix
\section{Additional Figures \& Halo Properties}
\label{apx:more_details}
In this Section we summarize halo properties, and show additional visual representations of the halos that are not included in the projections in them main paper. The properties of the Pop~III forming halos are found in Table \ref{tab:halos}. The halos that have a recollapse time but no metallicity of the second-generation stars, are halos which form stars after the SNe, but only after merging with other halos that contain stars. As we do not model feedback from second-generation stars, we cannot determine the metallicities of these stars.
\begin{deluxetable}{cccccccccc}
    \tablecaption{\label{tab:halos}Properties of the halos forming Pop~III stars}
    \tablecolumns{10}\colnumbers
    \tablehead{
         \colhead{halo} & \colhead{$z_\mathrm{col}$} & \colhead{$z_\mathrm{sn}$} & \colhead{$z_\mathrm{rec}$} & \colhead{$M_\mathrm{vir,col}$} & \colhead{$M_\mathrm{vir,rec}$} & \colhead{$t_\mathrm{rec}$} & \colhead{$M_\mathrm{dens}$} & \colhead{[Fe/H]} & \colhead{[Fe/H]$_\mathrm{halo}$}\\
        \colhead{---} & \colhead{---} & \colhead{---} & \colhead{---} & \colhead{$10^6\,\Ms$} & \colhead{$10^6\,\Ms$} & \colhead{Myr} & \colhead{$10^3\Ms$} & \colhead{---} & \colhead{---}}
        \startdata
        0 & 24.0 & 23.8 & 19.2 & 2.7 & 7.1 & 47.6 & 2.7 & -3.0 & -3.1\\
        1 & 22.3 & 22.1 & 15.1 & 1.1 & 8.5 & 112  & 5.8 & -6.0 & -2.9\\ 
        2 & 21.8 & 21.6 & 17.9 & 1.6 & 5.5 & 50.5 & 6.8 & -3.2 & -2.9\\
        3 & 21.8 & 21.6 & 13.6 & 1.5 & 12.7 & 150 & 9.0 & ---  & ---\\
        4 & 20.4 & 20.3 & ---  & 0.8 & --- & ---  & 0.15 & ---  & ---\\
        5 & 20.4 &20.2 & 17.5 & 1.2 & 5.4 &  38.4 & 12.5 & ---  & ---\\
        6 & 20.0 & 19.8 & 15.7 & 1.6 & 5.8  & 70.7& 11.5 & ---  & ---\\
        7 & 20.0 & 19.8 & 15.0 & 1.7 & 4.8  & 88.4& 0.61 & ---  & --- \\
        8 & 19.9 & 19.7 & 19.5 & 1.7 & 1.8  & 2.8 & 12.8 & -2.0 & -0.7\\
        9 & 19.8 & 19.6 & 14.5 & 2.0 & 12.1 & 98.7& 2.4 & -2.1 & -2.3 \\
        10 & 19.8 & 19.6 & 16.7 & 1.3 & 6.1 & 48.2& 6.4 & -2.6 & -2.5 \\
        11 & 19.7 & 19.5 & 16.5 & 2.4 & 8.2 & 49.8& 15.3 & -1.5 & -1.9 \\
        12 & 19.7 & 19.5 & 19.1 & 1.9 & 2.3 & 5.8 & 16.2 & -3.4 & -2.1 \\
        13 & 19.6 & 19.5 & 19.0 & 1.9 & 2.3 & 5.6 & 18.6 & -1.6 & -1.4 \\
     \enddata
     \tablecomments{The columns are: halo number(1), redshift of first star formation (collapse redshift, 2), redshift of the supernova (3), redshift of recollapse (4), estimated virial mass at collapse (5), estimated virial mass at recollapse (6), recollapse time (7), mass of dense gas in the halo shortly before the SN (8), metallicity of 2nd generation star (9), metallicity of halo at recollapse (10). Note that halo 4 does not recollapse, and that halos 3 and 7 accrete Pop~II stars from external halos before recollapsing, and we therefore cannot make reliable metallicity determinations for the star formation during recollapse in these halos. }
\end{deluxetable}

\begin{figure*}
    \includegraphics[width=\textwidth]{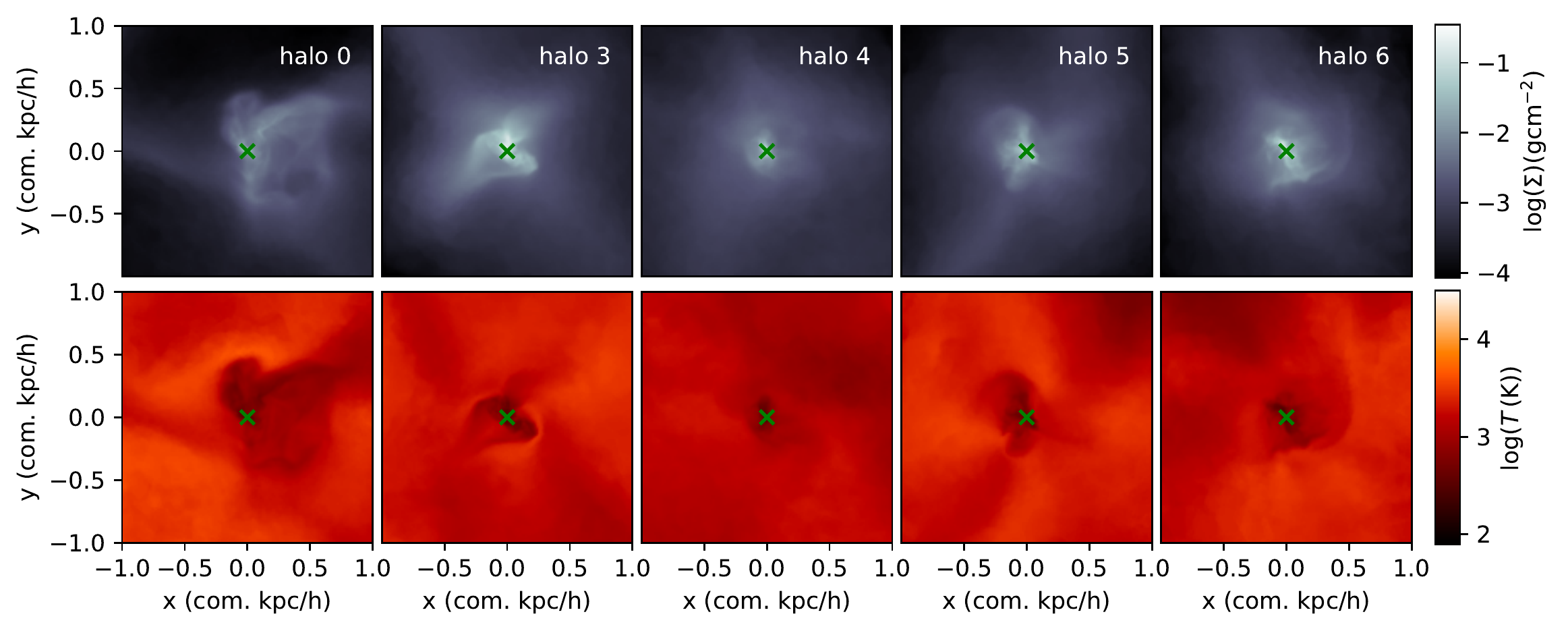}
    \caption{\label{fig:apx_sf1}As Fig. \ref{fig:SF}, but for halos 0, 3, 4, 5, and 6}
\end{figure*}

\begin{figure*}
    \includegraphics[width=\textwidth]{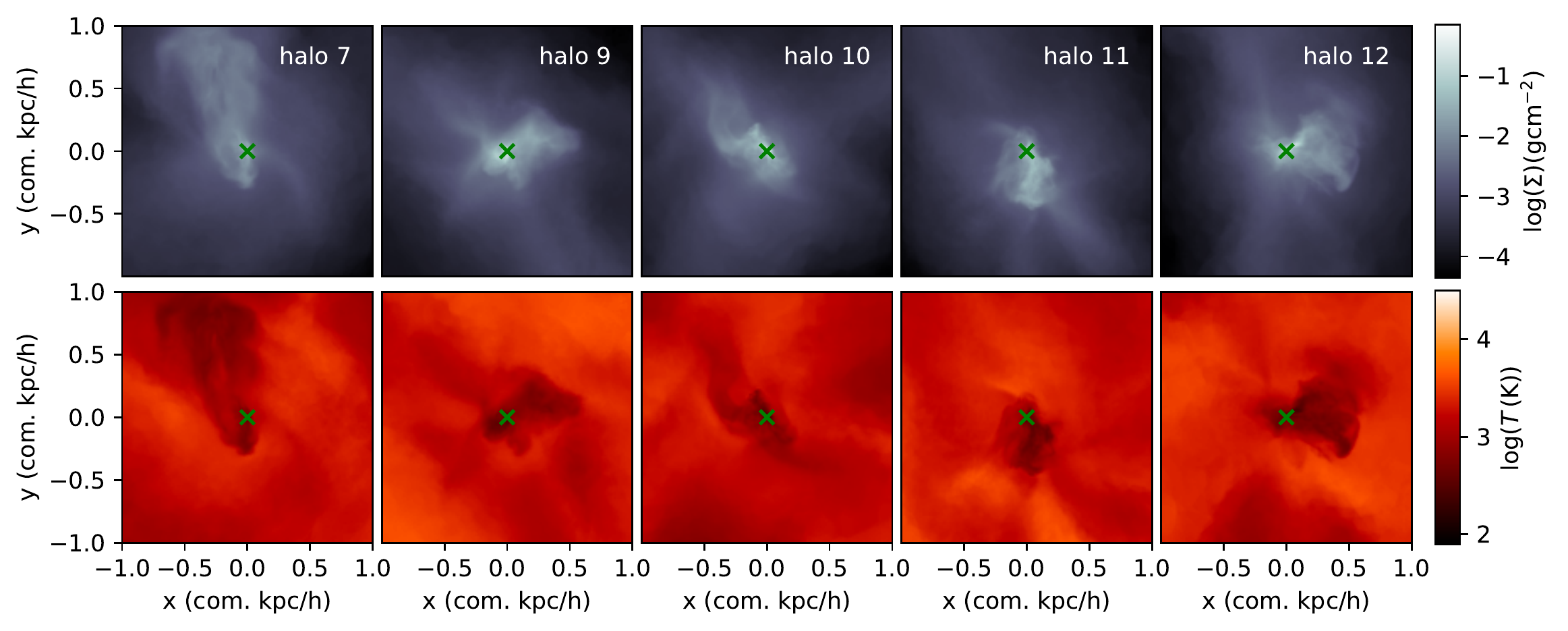}
    \caption{\label{fig:apx_sf2}As Fig. \ref{fig:SF}, but for halos 7, 9, 10, 11, and 12}
\end{figure*}

\begin{figure*}
    \includegraphics[width=\textwidth]{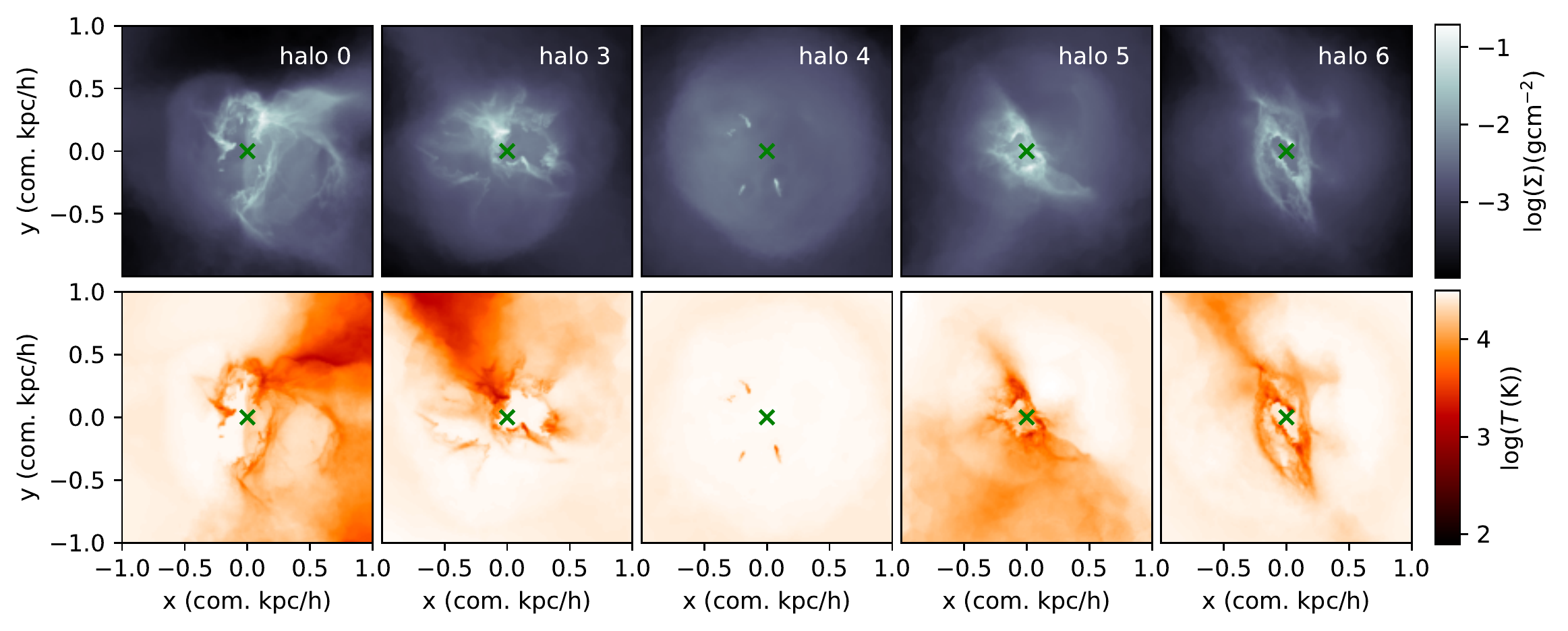}
    \caption{\label{fig:apx_sn1}As Fig. \ref{fig:pre_sn_imgs}, but for halos 0, 3, 4, 5, and 6}
\end{figure*}

\begin{figure*}
    \includegraphics[width=\textwidth]{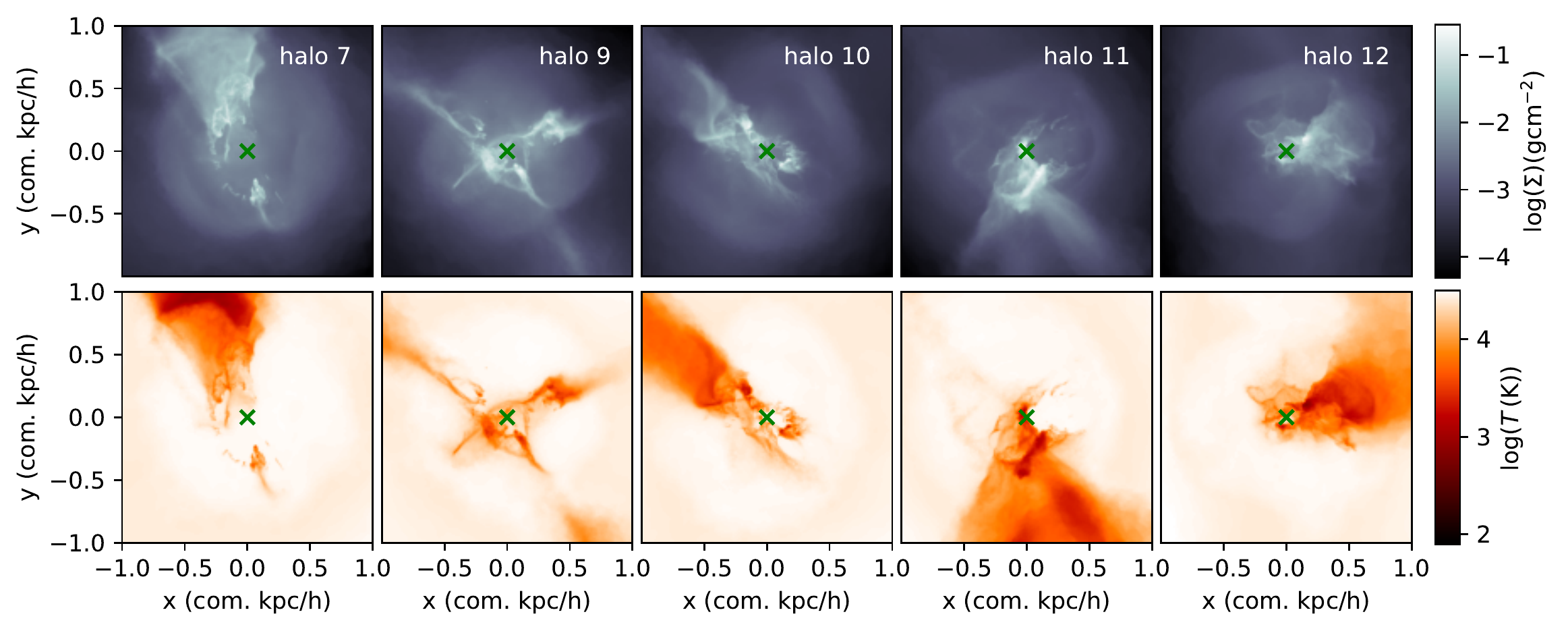}
    \caption{\label{fig:apx_sn2}As Fig. \ref{fig:pre_sn_imgs}, but for halos 7, 9, 10, 11, and 12}
\end{figure*}

\begin{figure*}
    \includegraphics[width=\textwidth]{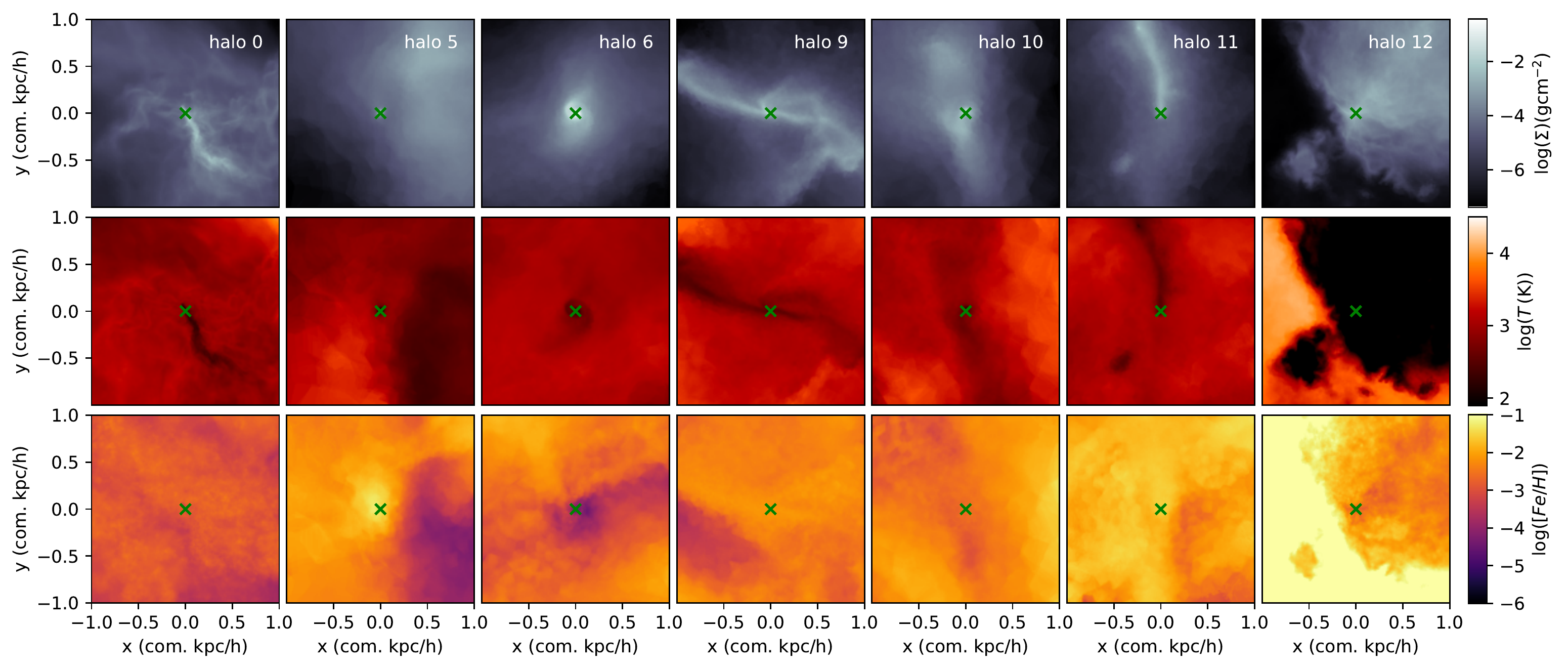}
    \caption{\label{fig:apx_rec}As Fig \ref{fig:rec} but for halos 0 5, 6, 9, 10, 11, and 12}
\end{figure*}

Halos 1, 2, 8, and 13 are shown in Figures \ref{fig:SF}, \ref{fig:pre_sn_imgs} and \ref{fig:rec} at the the time of the formation of the Pop~III star, shortly before the SN and at recollapse respectively. Similarily the remaining halos are shown in Figures \ref{fig:apx_sf1} and \ref{fig:apx_sf2} at star formation, in Figures \ref{fig:apx_sn1} and \ref{fig:apx_sn2} shortly before the SNe and in Fig. \ref{fig:apx_rec} at recollapse. We show these images for completeness and to give readers a visual impression of the halos. A more detailed analysis of the substructures and clumps seen within the figures will be subject to a follow-up study.

\begin{figure*}
    \includegraphics[width=\textwidth]{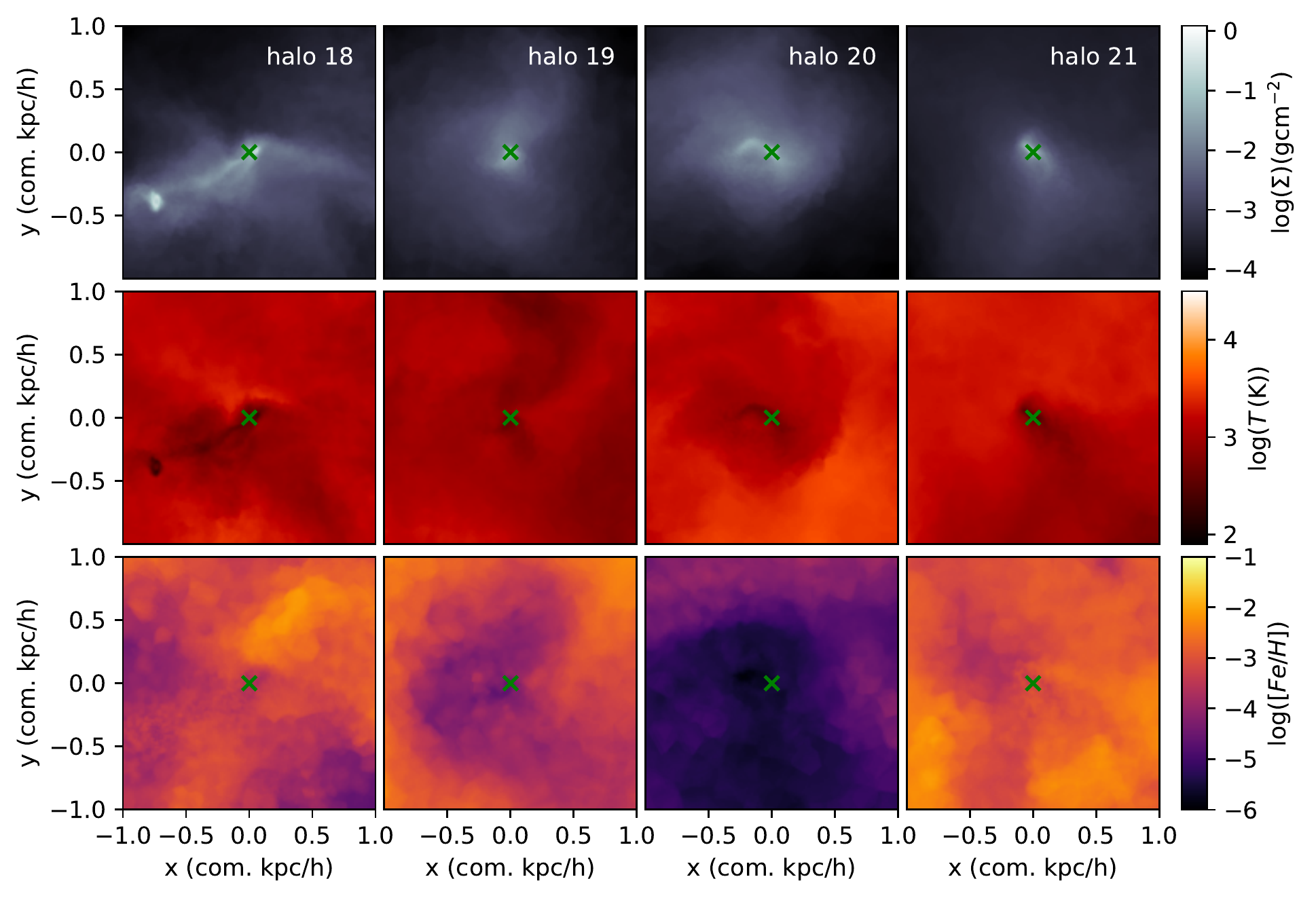}
    \caption{\label{fig:apx_ext}As Fig. \ref{fig:ext}, but for halos 18, 19, and 20}
\end{figure*}

The properties of the externally enriched halos which form stars with non-zero metallicity are summarized in Table \ref{tab:ext_halos}. The halos at star formation are show in Fig. \ref{fig:apx_ext} and the main paper in Fig. \ref{fig:ext}. 
\begin{deluxetable}{cccccccc}
    \tablecaption{Externally enriched halos\label{tab:ext_halos}}
    \tablecolumns{8}
    \tablehead{
         \colhead{halo} & \colhead{enriching halo} & \colhead{$z_\mathrm{col}$} & \colhead{$M_\mathrm{vir,col}$} & \colhead{$t_\mathrm{sn}$} & \colhead{d} & \colhead{[Fe/H]} & \colhead{[Fe/H]$_\mathrm{halo}$}\\
        \colhead{---} & \colhead{---} & \colhead{---} & \colhead{$10^6\,\Ms$} & \colhead{Myr} & \colhead{kpc/h} & \colhead{---} & \colhead{---}}
        \startdata
        14 & 18.2 & 12 & 0.9 &  19.8 &  7.3 & -4.6 & -3.4\\
        15 & 16.3 &  3 & 3.1 &  80.6 & 11.7 & -4.8 & -2.6\\
        16 & 15.8 & 12 & 2.0 &  66.2 & 25.2 & -6.0 & -3.3\\
        17 & 15.0 &  6 & 1.7 &  89.5 & 17.1 & -5.1 & -3.6\\
        18 & 14.5 &  2 & 0.7 & 123.1 & 19.0 & -4.5 & -2.9\\
        19 & 14.5 & 11 & 2.2 &  98.6 & 34.8 & -5.5 & -3.9\\
        20 & 14.1 &  1 & 0.9 & 138.8 &  5.4 & -3.3 & -2.8\\
        \enddata
    \tablecomments{The columns are: halo number(1), enriching halo (2), collapse redshift (3), estimated virial mass at collapse (4), time since SN (5), distance between star and the enriching SN (6), metallicity of the star (7), metallicity of halo at collapse (8).}
\end{deluxetable}

\bibliography{lit.bib}{}
\bibliographystyle{aasjournal}



\end{document}